\documentclass[10pt]{article}

   \usepackage{graphicx}
   \usepackage{amsfonts}
   \usepackage{amssymb}
   \usepackage{amsthm}
   \usepackage{amsmath}
   \usepackage{bbold}

   \usepackage[active]{srcltx}

   \usepackage{here}
   \usepackage{cite}
   \usepackage{color}
   \usepackage{soul}

   \usepackage[affil-it]{authblk}

   \usepackage{xcolor}
   \usepackage[
   bookmarks=true, linktocpage, colorlinks=true, allbordercolors=white, citecolor=blue]{hyperref}
   \usepackage{xspace}

   \graphicspath{{figs/}}

\newcommand{\be}{\begin{equation}}
\newcommand{\ee}{\end{equation}}

\newcommand{\beq}{\begin{equation}}
\newcommand{\eeq}{\end{equation}}

\newcommand\bea{\begin{eqnarray}}
\newcommand\eea{\end{eqnarray}}

\allowdisplaybreaks
\newcommand{\lsim}
{\;\raisebox{-.3em}{$\stackrel{\displaystyle <}{\sim}$}\;}
\newcommand{\gsim}
{\;\raisebox{-.3em}{$\stackrel{\displaystyle >}{\sim}$}\;}

\newcommand\al{\alpha}

\newcommand\tb{\tan\beta}

\newcommand\ReDiag{\mathop{%
  \raise .5pt\hbox{[}%
  \widetilde{\mathrm{Re}}%
  \raise .5pt\hbox{]}}}
\newcommand\ReOffDiag{\mathop{%
  \raise .5pt\hbox{$\llbracket$}%
  \widetilde{\mathrm{Re}}%
  \raise .5pt\hbox{$\rrbracket$}}}

\newcommand\DRbar{\ensuremath{\smash{\overline{\mathrm{DR}}}}}

\newcommand\Mh{M_h}

\newcommand\MA{M_A}

\newcommand\mb{m_b}
\newcommand\mt{m_t}

\newcommand\ino[1]{\tilde\chi_{#1}}

\newcommand\neu[1]{\ino{#1}^0}
\newcommand\mneu[1]{M_{\neu{#1}}}

\newcommand\refeq[1]{Eq.~(\ref{#1})}

\newcommand\refta[1]{Table~\ref{#1}}
\newcommand\refse[1]{Sect.~\ref{#1}}

\newcommand{\tev}{\,\, \mathrm{TeV}}
\newcommand{\gev}{\,\, \mathrm{GeV}}

\newcommand\FH{\texttt{FeynHiggs}\xspace}

\newcommand\ab{\ensuremath{\mbox{ab}}}

\newcommand\iab{\ensuremath{\ab^{-1}}}

\newcommand{\br}{\text{BR}}

\newcommand{\sig}{\sigma}

\def\reffi#1{\mbox{Fig.~\ref{#1}}}

\newcommand{\VL}{\left( \begin{array}{c}}
\newcommand{\VR}{\end{array} \right)}
\newcommand{\ML}{\left( \begin{array}{cc}}
\newcommand{\MLd}{\left( \begin{array}{ccc}}
\newcommand{\MLv}{\left( \begin{array}{cccc}}
\newcommand{\MR}{\end{array} \right)}

\definecolor{Lightblue}{cmyk}{0.9,0.1,0.1,0.3}
\definecolor{dgelborange}{cmyk}{0.,0.3,0.5, 0.}
\definecolor{Orange}{cmyk}{0.,0.5,0.5, 0.}
\definecolor{Lila}{rgb}{0.5,0.,1}

\newcommand{\htb}[1]{{\color{blue} #1}}

\newcommand{\complex}{{{\rm I} \kern -.59em {\rm C}}}

\begin{document}

\thispagestyle{empty}

\title{
\vspace*{-3.5cm}
Probing Unified Theories with Reduced Couplings \\ at  Future Hadron Colliders}
\date{}
\author{ \hspace*{7mm} S. Heinemeyer$^{1,2,3}$\thanks{email: Sven.Heinemeyer@cern.ch} ,
J. Kalinowski$^4$\thanks{email: kalino@fuw.edu.pl} ,
W. Kotlarski$^5$\thanks{email: wojciech.kotlarski@tu-dresden.de} ,
 M. Mondrag\'on$^6$\thanks{email: myriam@fisica.unam.mx} ,

G. Patellis$^7$\thanks{email: patellis@central.ntua.gr} ,
N. Tracas$^7$\thanks{email: ntrac@central.ntua.gr}$\,$
and G. Zoupanos$^{7,8,9}$\thanks{email: George.Zoupanos@cern.ch}\\
{\small
$^1$Instituto de F\'{\i}sica Te\'{o}rica (UAM/CSIC), Universidad Aut\'{o}noma de Madrid Cantoblanco, 28049 Madrid, Spain\\
$^2$Campus of International Excellence UAM+CSIC, Cantoblanco, 28049 Madrid, Spain \\
$^3$Instituto de F\'{\i}sica de Cantabria (CSIC-UC), E-39005 Santander, Spain \\
$^4$University of Warsaw - Faculty of Physics,
 ul.\ Pasteura 5, 02-093 Warsaw, Poland \\
$^5$Technische Universit\"at Dresden - Institut für Kern- und Teilchenphysik (IKTP)
01069 Dresden, Germany \\
$^6$Instituto de F\'{\i}sica, Universidad Nacional Aut\'onoma de M\'exico, A.P. 20-364, CDMX 01000 M\'exico\\ %
$^7$ Physics Department,   Nat. Technical University, 157 80 Zografou, Athens, Greece\\
$^8$ Max-Planck Institut f\"ur Physik, F\"ohringer Ring 6, D-80805
  M\"unchen, Germany \\
$^9$ Theoretical Physics Department, CERN, Geneva, Switzerland \\
}
}

\maketitle

\abstract{
The search for renormalization group invariant relations among parameters
to all orders in perturbation theory constitutes the basis of the
reduction of couplings concept. Reduction of couplings can be achieved in
certain $N=1$ supersymmetric Grand Unified Theories and few of them can
become even finite at all loops. We review the basic idea, the tools that
have been developed as well as the resulting  theories in which successful
reduction of couplings has been achieved so far. These include: (i) a
reduced version of the minimal $N = 1$ $SU(5)$ model, (ii) an all-loop finite
$N = 1$ $SU(5)$ model, (iii) a two-loop finite $N = 1$ $SU(3)^3$ model and
finally (vi) a reduced version of the Minimal Supersymmetric Standard
Model. In this paper we present a number of benchmark scenarios for each model and 
investigate their observability at existing and future hadron colliders. 
The heavy supersymmetric spectra featured by each of the above models
are found to be beyond the reach of the 14 TeV HL-LHC. It is also found that the reduced 
version of the MSSM is already ruled out by the LHC
 searches for heavy neutral MSSM Higgs bosons.  In turn the
discovery potential of the 100~TeV FCC-hh is investigated and 
found that 
large parts of the predicted spectrum of these models can be tested,
but the higher mass regions are beyond the reach even of the FCC-hh.\\
}

\vspace{1.5cm}

\noindent
{\tt
IFT-UAM/CSIC-20-152
}

\newpage
\pagebreak

	\section{Introduction}
The \textit{reduction of~couplings} method
\cite{book,Zimmermann:1984sx,Oehme:1984yy,Oehme:1985jy}
(see also \cite{Ma:1977hf,Chang:1974bv,Nandi:1978fw})
is a~promising method which relates~originally seemingly independent parameters to a single, ``primary'' coupling.
The~method requires the original theory to which it is applied to be a renormalizable~one,
and the resulting relation among the parameters~to be valid at all energy scales, i.e.\
Renormalization Group Invariant (RGI).

A next (natural) step,~after the introduction of a novel symmetry through a Grand Unified Theory (GUT)
\cite{Pati:1973rp,Georgi:1974sy,Georgi:1974yf,Fritzsch:1974nn,Gursey:1975ki,Achiman:1978vg}),
in order to achieve reduction~of free parameters of the SM is the relation of the gauge sector  to the   Yukawa sector
(Gauge Yukawa Unification, GYU).~This was the central characteristic of the \textit{reduction of couplings} approach in the first period of searches, applied in $N=1$ GUTs \cite{Kapetanakis:1992vx,Mondragon:1993tw,Kubo:1994bj,Kubo:1994xa,Kubo:1994sf,Kubo:1995xu,Kubo:1995cg,Kubo:1995hm,Kubo:1995zg,Kubo:1996en,Kubo:1997fi,Kobayashi:1997qx,Kubo:1997rn,Mondragon:2009zz}.
According to that approach, being~in a GUT environment, RGI relations are set between the unification scale and the Planck scale.
One-loop consideration can guarantee~the all-loop validity of those relations. Moreover, RGI relations can be found
which guarantee all order finiteness~of a theory.
The method has predicted the top quark mass~in the finite $N=1$ $SU(5)$ model
\cite{Kapetanakis:1992vx,Mondragon:1993tw}
as well as in~the minimal $N=1$ $SU(5)$ one
\cite{Kubo:1994bj}
before its experimental measurement
\cite{Lancaster:2011wr}.

Since SuperSymmetry~(SUSY) seems an essential ingredient for the \textit{reduction of couplings} method,
we have to include a~supersymmetry breaking sector (SSB), which involves dimension-1 and -2 couplings.
The supergraph method~and the spurion superfield technique played an important role for the progress in that sector,
leading to complete all-loop~finite models, i.e. including the SSB sector.
The all-loop finite $N=1$ $SU(5)$ model
\cite{Heinemeyer:2007tz}
has given~a prediction for the Higgs~mass compatible with the experimental results
\cite{Aad:2012tfa,Chatrchyan:2012ufa,ATLAS:2013mma}
and a heavy SUSY mass spectrum, consistent with the experimental non-observation of these particles.
The reduction of couplings method has been applied to several other cases.
The full analysis of the most successful models, that
includes~predictions in agreement with the experimental measurements of
the top and bottom quark masses for each~model, can be found in a
recent work \cite{Heinemeyer:2020ftk}. 

In this paper we address the question to what extent the reduction of couplings idea, as applied in the so far phenomenologically successful models,  can be experimentally tested at HL-LHC and future FCC hadron
 collider. To this end we propose a number of benchmark points for each model.  We present the 
 SUSY breaking parameters used as input  in each benchmark to calculate the corresponding Higgs boson and supersymmetric particles masses.  Then we compute the expected production cross sections at the 14 TeV (HL-)LHC and the 100 TeV FCC-hh and investigate which production channels can be observed.  

The present work is~organized as follows. In Section 2 we review the basic idea of the reduction of couplings.~In Section 3 we list the phenomenological constraints used in our analyses, wile in Section 4 we explain the computational setup. In Sections 5, 6, 7 and 8 we review four interesting  models, namely
(i) the Minimal $N=1$ Supersymmetric $SU(5)$,
(ii) the Finite $N=1$ Supersymmetric $SU(5)$,
(iii) the Finite $SU(3)^3$ and
(iv) the MSSM, in which the reduction of couplings has been successfully applied.
We briefly review some earlier results of our phenomenological analysis. In
this context the new version of the
{\tt FeynHiggs}~\cite{Degrassi:2002fi,BHHW,FeynHiggs,Bahl:2019hmm} code
plays a crucial role, which was used to calculate the Higgs-boson
predictions, in  particular the mass of the lightest CP-even Higgs boson.
The improved predictions of {\tt FeynHiggs} are compared with the LHC
measurements and the Beyond Standard Model (BSM) Higgs boson searches.
As a new part of the analysis we examine in each model the discovery
potential of the Higgs and SUSY spectrum at approved future and
hypothetical future hadron colliders. ~Finally, Section 8 is dedicated
to brief conclusive remarks.

%%%%%%%%%%%%%%%%%%%%%%%%%%%%%%%%%%%%%%%%%
\section{Theoretical Basis}\label{sec2}
Here we will briefly~review the core idea of the \textit{reduction of couplings} method. The target is to single out a~basic
parameter (which we will call the primary coupling), where all other parameters can be expressed in terms of this~one
through RGI relations. Such a relation has, in general,  the~form $\Phi (g_1,\cdots,g_A) ~=~\mbox{const.}$
which should satisfy the following partial differential~equation (PDE)
\beq
\mu\,\frac{d \Phi}{d \mu} = {\vec \nabla}\Phi\cdot {\vec \beta} ~=~
\sum_{a=1}^{A}
\,\beta_{a}\,\frac{\partial \Phi}{\partial g_{a}}~=~0~,
\eeq
where $\beta_a$ is~the  $\beta$-functions of $g_a$.
The above PDE is~equivalent to the following set of ordinary differential equations (ODEs), which are called Reduction~Equations (REs)
\cite{Zimmermann:1984sx,Oehme:1984yy,Oehme:1985jy},
\beq
\beta_{g} \,\frac{d g_{a}}{d g} =\beta_{a}~,~a=1,\cdots,A-1~,
\label{redeq}
\eeq
where now $g$ and $\beta_g$ are~the primary coupling and its corresponding $\beta$-function.
There are obviously $A-1$ relations~in the form of $\Phi (g_1,\cdots,g_A)~=~\mbox{const.}$ in order to
express all other couplings in term~of the primary one.

The crucial demand is that the above~REs admit power series solutions
\beq
g_{a} = \sum_{n}\rho_{a}^{(n)}\,g^{2n+1}~,
\label{powerser}
\eeq
which preserve perturbative~renormalizability. Without this requirement, we just
trade each ``dependent'' coupling~for an integration constant. The power series,
which are a set of special~solutions, fix that constant. It is very important to point out that
the uniqueness of such a~solution can be already decided at the one-loop level
\cite{Zimmermann:1984sx,Oehme:1984yy,Oehme:1985jy}.
In supersymmetric theories,~where the asymptotic behaviour of several parameters are similar,
the use of power series as~solutions of the REs are justified. But, usually, the reduction
is not ``complete'', which~means that not all of the couplings can be reduced in favor
of the primary one, leading~to the so called ``partial reduction''
\cite{Kubo:1985up,Kubo:1988zu}.

We proceed to the reduction~scheme for massive parameters, which is far from being straightforward.
A  number of conditions is~required (see for example
\cite{Piguet:1989pc}).
Nevertheless, progress has~been achieved, starting from~\cite{Kubo:1996js}, and finally we can introduce
mass parameters and couplings~carrying mass dimension
\cite{Breitenlohner:2001pp,Zimmermann:2001pq}
in the same way as dimensionless~couplings.

Consider the~superpotential
\beq
W= \frac{1}{2}\,\mu^{ij} \,\Phi_{i}\,\Phi_{j}+
\frac{1}{6}\,C^{ijk} \,\Phi_{i}\,\Phi_{j}\,\Phi_{k}~,
\label{supot}
\eeq
and the SSB~sector Lagrangian
\beq
-{\cal L}_{\rm SSB} =
\frac{1}{6} \,h^{ijk}\,\phi_i \phi_j \phi_k
+
\frac{1}{2} \,b^{ij}\,\phi_i \phi_j
+
\frac{1}{2} \,(m^2)^{j}_{i}\,\phi^{*\,i} \phi_j+
\frac{1}{2} \,M\,\lambda_i \lambda_i+\mbox{h.c.},
\label{supot_l}
\eeq
where $\phi_i$'s are the scalar~fields of the corresponding superfields $\Phi_i$'s and
$\lambda_i$ are the gauginos.

Let us write down some well~known relations:\\
(i) The $\beta$-function of~the gauge coupling at one-loop level is given by
\cite{Parkes:1984dh,West:1984dg,Jones:1985ay,Jones:1984cx,Parkes:1985hh}
\beq
\beta^{(1)}_{g}=\frac{d g}{d t} =
  \frac{g^3}{16\pi^2}\,\left[\,\sum_{i}\,T(R_{i})-3\,C_{2}(G)\,\right]~,
\label{betag}
\eeq
where $T(R_i)$ is the Dynkin~index of the  rep $R_i$ where the matter fields belong and $C_2(G)$ is the quadratic Casimir operator~of the adjoint rep $G$.\\
(ii) The anomalous~dimension $\gamma^{(1)}\,^i_j$, at a one-loop level, of a chiral~superfield  is
\beq
\gamma^{(1)}\,^i_j=\frac{1}{32\pi^2}\,\left[\,
C^{ikl}\,C_{jkl}-2\,g^2\,C_{2}(R_{i})\delta^i_j\,\right]~.
\label{gamay}
\eeq
(iii) The $\beta$-functions~of $C_{ijk}$'s, at one-loop level, following the
$N = 1$ non-renormalization~theorem
\cite{Wess:1973kz,Iliopoulos:1974zv,Fujikawa:1974ay},
are expressed in terms of~the anomalous dimensions of the fields involved
\beq
\beta_C^{ijk} =
  \frac{d C_{ijk}}{d t}~=~C_{ijl}\,\gamma^{l}_{k}+
  C_{ikl}\,\gamma^{l}_{j}+
  C_{jkl}\,\gamma^{l}_{i}~.
\label{betay}
\eeq
We proceed by assuming
that the REs admit~power~series~solutions:
\beq
C^{ijk} = g\,\sum_{n=0}\,\rho^{ijk}_{(n)} g^{2n}~.
\label{Yg}
\eeq
Trying to obtain all-loop~results we turn to relations among $\beta$-functions.
The spurion technique
\cite{Fujikawa:1974ay,Delbourgo:1974jg,Salam:1974pp,Grisaru:1979wc,Girardello:1981wz}
gives all-loop relations~among SSB $\beta$-functions
\cite{Yamada:1994id,Kazakov:1997nf,Jack:1997pa,Hisano:1997ua,Jack:1997eh,Avdeev:1997vx,Kazakov:1998uj}.
Then, assuming that the~reduction of $C^{ijk}$ is possible to all orders
\beq
\label{Cbeta}
\frac{dC^{ijk}}{dg} = \frac{\beta^{ijk}_C}{\beta_g}~,
\eeq
as well as for $h^{ijk}$
\beq
\label{h2NEW}
h^{ijk} = - M \frac{dC^{ijk}}{d\ln g}~,
\eeq
it can be~proven
\cite{Jack:1999aj,Kobayashi:1998iaa}
that the following~relations are all-loop RGI
\begin{align}
M &= M_0~\frac{\beta_g}{g} ,  \label{M-M0} \\
h^{ijk}&=-M_0~\beta_C^{ijk},  \label{hbeta}  \\
b^{ij}&=-M_0~\beta_{\mu}^{ij},\label{bij}\\
(m^2)^i_j&= \frac{1}{2}~|M_0|^2~\mu\frac{d\gamma^i{}_j}{d\mu},
\label{scalmass}
\end{align}
where $M_0$~is an~arbitrary~reference mass scale to be specified and \refeq{M-M0} is the Hisano-Shifman relation \cite{Hisano:1997ua}
(note that in both assumptions~we do not rely on specific solutions of these equations).\\
As a next step we substitute~the last equation, Eq.~(\ref{scalmass}), by a more general RGI sum rule that
holds to~all orders \cite{Kobayashi:1998jq}
\begin{equation}
\begin{split}
m^2_i+m^2_j+m^2_k &=
|M|^2 \left\{~
\frac{1}{1-g^2 C_2(G)/(8\pi^2)}\frac{d \ln C^{ijk}}{d \ln g}
+\frac{1}{2}\frac{d^2 \ln C^{ijk}}{d (\ln g)^2}~\right\}\\
& \qquad\qquad +\sum_l
\frac{m^2_l T(R_l)}{C_2(G)-8\pi^2/g^2}
\frac{d \ln C^{ijk}}{d \ln g}~,
\label{sum2}
\end{split}
\end{equation}
which leads~to the following one-loop relation
\beq
m^2_i+m^2_j+m^2_k=|M|^2~.\label{1loopsumrule}
\eeq
Finally, note~that in the case of product gauge groups,  Eq.~(\ref{M-M0}) takes the form
\beq
M_i=\frac{\beta_{g_i}}{g_i}M_0~,
\eeq
where $i$ denotes~the group of the product. This will be used in the Reduced MSSM case.

Consider an $N=1$ globally~supersymmetric gauge theory, which is chiral and anomaly free,
where $G$ is the gauge group~and $g$ the associated gauge coupling. The theory has the
superpotential of Eq.~(\ref{supot}), while~the one-loop gauge and $C_{ijk}$s $\beta$-functions
are given by Eq.~(\ref{betag}) and~Eq.~(\ref{betay}) respectively and the one-loop anomalous dimensions
of the chiral superfields~by Eq.~(\ref{gamay}).\\
Demanding the vanishing of~all one-loop $\beta$-functions, Eqs.(\ref{betag},\ref{gamay}) lead to
the relations
\begin{align}
\sum _i T(R_{i})& = 3 C_2(G) \,,
\label{1st}     \\
 C^{ikl} C_{jkl} &= 2\delta ^i_j g^2  C_2(R_i)~.
\label{2nd}
\end{align}
The finiteness conditions~for an $N=1$ supersymmetric theory with $SU(N)$
associated group is found~in \cite{Rajpoot:1984zq} while discussion of the
no-charge renormalization~and anomaly free requirements can be found in  \cite{Rajpoot:1985aq}.
It should be noted that~conditions (\ref{1st}) and (\ref{2nd}) are necessary and sufficient to
ensure finiteness at the~two-loop level
\cite{Parkes:1984dh,West:1984dg,Jones:1985ay,Jones:1984cx,Parkes:1985hh}.

The requirement of~finiteness, at the one-loop level, in softly broken SUSY theories
demands additional~constraints among the soft terms of the SSB sector \cite{Jones:1984cu},
while, once more,~these one-loop requirements assure two-loop finiteness, too \cite{Jack:1994kd}.
These conditions~impose restrictions on the irreducible~representations $R_i$ of the gauge group
$G$ as well as on~the Yukawa couplings. For example, since $U(1)$s are not compatible with condition
(\ref{1st}), the~MSSM is excluded. Therefore, a GUT is initially required with the MSSM being its low energy theory.
Also, since~condition (\ref{2nd}) forbids the appearance of gauge singlets ($C_2(1)=0$),
F-type~spontaneous symmetry~breaking \cite{O'Raifeartaigh:1975pr} are not compatible with finiteness.
Finally, D-type spontaneous~breaking \cite{Fayet:1974jb} is also incompatible since it requires a
$U(1)$ group.

The nontrivial point is~that the relations among couplings (gauge and Yukawa) which are imposed
by the~conditions (\ref{1st}) and (\ref{2nd}) should hold at any energy scale.
The necessary~and sufficient~condition is~to
require that such~relations are~solutions to~the  REs (see \refeq{Cbeta})
\beq
\beta _g
\frac{d C_{ijk}}{dg} = \beta _{ijk}
\label{redeq2}
\eeq
holding at all~orders. We note, once more, that the existence of one-loop level power series solution
guarantees the~all-order series.

There exist the~following theorem
\cite{Lucchesi:1987he,Lucchesi:1987ef}
which points~down which are the  necessary and sufficient conditions
in order for~an $N=1$ SUSY theory to be all-loop finite.
In refs
\cite{Lucchesi:1987he,Lucchesi:1987ef, Piguet:1986td,Piguet:1986pk,Ensign:1987wy,Lucchesi:1996ir,Piguet:1996mx} it was~shown that for an $N=1$ SUSY Yang-Mills~theory, based on a simple gauge group,
if the following~four conditions are fulfilled:\\
(i) No gauge~anomaly is present.\\
(ii) The $\beta$-function of~the gauge coupling is zero at one-loop level
\beq
\beta^{(1)}_g = 0 =\sum_i T(R_{i})-3\,C_{2}(G).
\eeq
(iii) The condition of~vanishing for the one-loop anomalous dimensions of matter fields,
\beq
  \gamma^{(1)}{}_{j}^{i}~=~0
  =\frac{1}{32\pi^2}~[ ~
  C^{ikl}\,C_{jkl}-2~g^2~C_{2}(R)\delta_j^i ],
\eeq
admits solution~of the form
\beq
C_{ijk}=\rho_{ijk}g,~\qquad \rho_{ijk}\in \mathbb{C}~.
\label{soltheo}
\eeq
(iv) When considered~as solutions of vanishing Yukawa~$\beta$-functions
(at one-loop order),~i.e. $\beta_{ijk}=0$, the above solutions are isolated and non-degenerate;\\
then,~each of the~solutions in Eq.~(\ref{soltheo})~can be extended uniquely to a formal power series in $g$,
and the~associated~super Yang-Mills models~depend on the~single coupling constant $g$ with a vanishing, at all orders, $\beta$-function.

While the validity~of the above cannot be extended to non-SUSY theories, it should be noted that reduction of couplings~and
finiteness are~intimately related.

%%%%%%%%%%%%%%%%%%%%%%%%%%%%%%%%%%%%%%%%%
\section{Phenomenological Constraints}\label{se:constraints}

In this section~we~briefly review several experimental constraints that were applied
in our phenomenological~analysis. The used values do not correspond to the latest experimental results, which, however, has a~negligible impact on our analysis.

In our models we~evaluate the pole mass of the top quark while the bottom quark mass
is evaluated at~the $M_Z$ scale (to avoid uncertainties to its pole mass). The experimental values, taken from
ref.\cite{Tanabashi:2018oca} are:
\beq
m_t^{\rm exp} = 173.1 \pm 0.9 \gev ~~~~~~,~~~~~~ m_b(M_Z) = 2.83 \pm 0.10 \gev~.
\label{mtmbexp}
\eeq
We interpret the~Higgs-like particle discovered in July 2012 by ATLAS and CMS
\cite{Aad:2012tfa,Chatrchyan:2012ufa}
as the~light CP-even Higgs~boson of~the MSSM \cite{Mh125,hifi,hifi2}.
The Higgs boson experimental average mass is \cite{Tanabashi:2018oca}
\footnote{This is the~latest available LHC combination. More recent
  measurements confirm~this value.}
\beq
M_h^{\rm exp}=125.10\pm 0.14~{\rm GeV}~.\label{higgsexpval}
\eeq
The theoretical~uncertainty \cite{Degrassi:2002fi,BHHW}, however, for the
prediction~of $M_h$ in~the MSSM dominates the total uncertainty, since it is much larger than the experimental~one.
In our following analyses we shall use the~new
{\tt FeynHiggs} code~\cite{Degrassi:2002fi,BHHW,FeynHiggs} (Version 2.16.0)
to predict the Higgs~mass.\footnote{An analysis of the impact of the improved $M_h$ calculation in various SUSY models can be found in \cite{Bagnaschi:2018igf}.}
{\tt FeynHiggs} evaluates~the Higgs masses based on a combination of
fixed~order~diagrammatic calculations and  resummation of
the (sub)leading~logarithmic contributions at all orders.
This provides a~reliable $M_h$ even for a large SUSY scale.
This new version~gives a downward shift on the Higgs mass $M_h$
of ${\cal O}(2~{\rm GeV})$ for~large SUSY masses and in particular gives a reliable point-by-point evaluation of the Higgs-boson mass uncertainty \cite{Bahl:2019hmm}. The theoretical uncertainty calculated
is added linearly to~the experimental error in Eq.~(\ref{higgsexpval}).

Furthermore, recent results from the ATLAS experiment \cite{Aad:2020zxo} set limits to the mass of the pseudoscalar Higgs boson, $M_A$, in comparison with $\tan{\beta}$. For models with $\tan{\beta}\sim 45-55$, as the ones examined here, the lowest limit for the physical pseudoscalar Higgs mass is 
\beq M_A\gtrsim  1900 {\rm ~GeV}.\eeq

We also consider the~following four flavor observables where SUSY has non-negligible impact.
For the~branching ratio $\br(b \to s \gamma)$ we~take a~value
from \cite{bsgth,HFAG}, while~for~the~branching~ratio $\br(B_s \to \mu^+ \mu^-)$ we~use a~combination of
\cite{Bobeth:2013uxa,RmmMFV,Aaij:2012nna,CMSBsmm,BsmmComb}:
\beq
\frac{\br(b \to s \gamma )^{\rm exp}}{\br(b \to s \gamma )^{\rm SM}} = 1.089 \pm 0.27~~~~~,~~~~~
\br(B_s \to \mu^+ \mu^-) = (2.9\pm1.4) \times 10^{-9}~.
\eeq
For~the $B_u$ decay~to $\tau\nu$ we use \cite{SuFla,HFAG,PDG14} and~for $\Delta M_{B_s}$  we~use \cite{Buras:2000qz,Aaij:2013mpa}:
\beq
\frac{\br(B_u\to\tau\nu)^{\rm exp}}{\br(B_u\to\tau\nu)^{\rm SM}}=1.39\pm 0.69~~~~~~~~~,~~~~~~~~~
\frac{\Delta M_{B_s}^{\rm exp}}{\Delta M_{B_s}^{\rm SM}}=0.97\pm 0.2~.
\eeq

Finally, we consider Cold Dark Matter (CDM) constraints.
Since the Lightest
SUSY Particle (LSP), which in our case is the lightest neutralino, is a
promising CDM candidate \cite{EHNOS}, we examine if each model is within the CDM relic density experimental limits.
The current bound~on the CDM relic density at
$2\,\sigma$~level is~given by \cite{Aghanim:2018eyx}%

\beq
\Omega_{\rm CDM} h^2 = 0.1120 \pm 0.0112~.
\label{cdmexp}
\eeq
In the~following sections~we will apply these~constraints to each model and discuss the corresponding collider~phenomenology.

\section{Computational setup}
\begin{figure}
\centering
\includegraphics[width=0.8\textwidth]{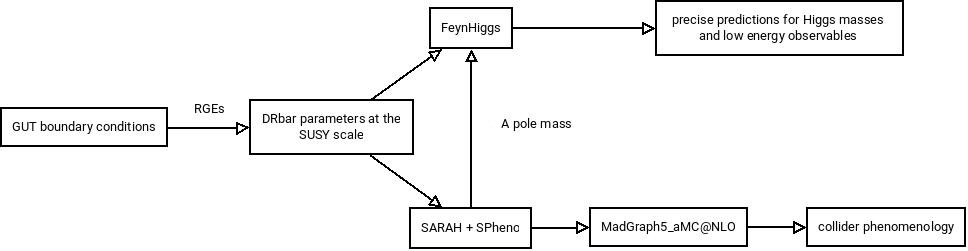}
\caption{Flow of information between used computer codes (see text for details).\label{fig:flow}}
\end{figure}
The setup for our phenomenological analysis is as follows.
Starting from an appropriate set of MSSM boundary conditions at the GUT scale,
parameters are run down to the SUSY scale using a private code. Two-loop RGEs are used throughout, with the exception of the soft sector, in which one-loop RGEs are used.
The running parameters are then used as inputs for both \FH
\cite{Degrassi:2002fi,BHHW,FeynHiggs,Bahl:2019hmm} and a
\texttt{SARAH}~\cite{Staub:2013tta} generated, custom MSSM module for
\texttt{SPheno}~\cite{Porod:2003um,Porod:2011nf}. It should be noted
that \FH requires the $m_b(m_b)$ scale, the \textit{physical} top quark
mass $m_t$ as well as {\it the physical} pseudoscalar boson mass $\MA$
as input.  The first two  values are calculated by the private code
while $\MA$ is calculated only in \DRbar~scheme. 
This single value is obtained from the \texttt{SPheno} output where it
is calculated at the two-loop level in the gaugeless limit
\cite{Gabelmann:2018axh,Goodsell:2015ira}.  The flow of information
between codes in our analysis is summarised in Fig.~\ref{fig:flow}. 

At this point both codes contain  a consistent set of all required parameters.
SM-like Higgs boson mass as well as low energy observables mentioned in Sec.~\ref{se:constraints} are evaluated using \FH.
To obtain collider predictions we use \texttt{SARAH} to generate \texttt{UFO}~\cite{Degrande:2011ua,Staub:2012pb} model for \texttt{MadGraph} event generator. 
Based on SLHA spectrum files generated by \texttt{SPheno},
we use \texttt{MadGraph5\_aMC@NLO}~\cite{Alwall:2014hca} to calculate cross sections for Higgs boson and SUSY particle production at the HL-LHC and a 100 TeV FCC-hh. 
Processes are generated at the leading order, using
\texttt{NNPDF31\_lo\_as\_0130} \cite{Ball:2017nwa}  structure functions interfaced through \texttt{LHAPDF6}~\cite{Buckley:2014ana}.
Cross sections are computed using dynamic scale choice, where the scale is set equal to the transverse mass of an event, in 4 or 5-flavor scheme depending on the presence or not of $b$-quarks in the final state. The results are 
given in Sec.~\ref{sec:minimalsu5}, \ref{sec:finitesu5} and \ref{sec:su33}.

%%%%%%%%%%%%%%%%%%%%%%%%%%%%%%%%%%%%%%%%%%%%%%%%%%%%%%%%%%%%%%%%%%%%%%%%%%%%%%%%%%%%%%%%%%%%%%%%%%%%%%
\section{The Minimal $N=1$ Supersymmetric $SU(5)$ Model}\label{sec:minimalsu5}
We start with the partial~reduction of the $N=1$ SUSY $SU(5)$ model
\cite{Kubo:1994bj,Kubo:1996js}. Our~notation is as follows:
$\Psi^{I}({\bf 10})$ and~$\Phi^{I}(\overline{\bf 5})$ refer to the three generations
of leptons and~quarks ($I=1,2,3$),
 $\Sigma({\bf 24})$ is the~adjoint which breaks $SU(5)$ to $SU(3)_{\rm C}\times SU(2)_{\rm L} \times U(1)_{\rm Y}$
and $\overline{H}({\overline{\bf 5}})$ represent the two~Higgs superfields for the~electroweak~symmetry breaking (ESB)~\cite{dimop,sakai}.
The choice of~using only one set of $({\bf 5} + {\bf \bar{5}})$ for the ESB renders the model asymptotically~free
(i.e. $\beta_g<0$ ). The~superpotential of the model is~described by
\beq
\begin{split}
W &= \frac{g_{t}}{4}\,
\epsilon^{\alpha\beta\gamma\delta\tau}\,
\Psi^{(3)}_{\alpha\beta}\Psi^{(3)}_{\gamma\delta}H_{\tau}+
\sqrt{2}g_b\,\Phi^{(3) \alpha}
\Psi^{(3)}_{\alpha\beta}\overline{H}^{\beta}+
\frac{g_{\lambda}}{3}\,\Sigma_{\alpha}^{\beta}
\Sigma_{\beta}^{\gamma}\Sigma_{\gamma}^{\alpha}+
g_{f}\,\overline{H}^{\alpha}\Sigma_{\alpha}^{\beta} H_{\beta}\\
&+ \frac{\mu_{\Sigma}}{2}\,
\Sigma_{\alpha}^{\gamma}\Sigma_{\gamma}^{\alpha}+
\mu_{H}\,\overline{H}^{\alpha} H_{\alpha}~,
\end{split}
\eeq
where only~the third generation Yukawa couplings are taken~into account.
The indices~$\alpha,\beta,\gamma,\delta,\tau$ are $SU(5)$ ones.
A detailed~presentation~of the model can be~found in  \cite{mondragon1} as well as in \cite{polonsky1,Kazakov:1995cy}.

Our primary~coupling is the gauge coupling $g$. In this model the gauge-Yukawa unification~can be~achieved through
two~sets of solutions which are asymptotically~free \cite{mondragon1}:
\beq
\label{two_sol}
\begin{split}
a & : g_t=\sqrt{\frac{2533}{2605}} g + \mathcal{O}(g^3)~,~
g_b=\sqrt{\frac{1491}{2605}} g + \mathcal{O}(g^3)~,~
g_{\lambda}=0~,~
g_f=\sqrt{\frac{560}{521}} g + \mathcal{O}(g^3)~,\\
b & : g_t=\sqrt{\frac{89}{65}} g + \mathcal{O}(g^3)~,~
g_b=\sqrt{\frac{63}{65}} g + \mathcal{O}(g^3)~,~
g_{\lambda}=0~,~g_f=0~,
\end{split}
\eeq
where the~higher order~terms denote~uniquely computable
power series~in~$g$. Let us~note that the reduction of the~dimensionless sector
is independent of the dimensionful one.
These~solutions describe~the boundaries~of a RGI surface in the parameter~space which is AF and
where $g_f$~and $g_{\lambda}$ could be different from zero.
Therefore,~a partial reduction is possible~where
$g_{\lambda}$ and $g_f$ are independent~(non-vanishing) parameters without endangering asymptotic freedom (AF).
The proton decay~constraints favor~solution $a$, therefore we choose this one for our discussion.
\footnote{
$ g_{\lambda}=0 $ is~inconsistent, but $g_{\lambda}  \lsim 0.005$
is necessary~in order~for the proton
decay~constraint \cite{Kubo:1995cg} to be~satisfied.
A small $g_{\lambda} $ is~expected to not~affect the prediction of unification of  SSB~parameters.}

The SSB~Lagrangian~is
\beq
\begin{split}
-{\cal L}_{\rm soft} &=
m_{H_u}^{2}{\hat H}^{* \alpha}{\hat H}_{\alpha}
+m_{H_d}^{2}
\hat{\overline {H}}^{*}_{\alpha}\hat{\overline {H}}^{\alpha}
+m_{\Sigma}^{2}{\hat \Sigma}^{\dag~\alpha}_{\beta}
{\hat \Sigma}_{\alpha}^{\beta}
+\sum_{I=1,2,3}\,[\,
m_{\Phi^I}^{2}{\hat \Phi}^{* ~(I)}_{\alpha}{\hat \Phi}^{(I)\alpha}\\
& +\,m_{\Psi^I}^{2}{\hat \Psi}^{\dag~(I)\alpha\beta}
{\hat \Psi}^{(I)}_{\beta\alpha}\,]
+\{ \,
 \frac{1}{2}M\lambda \lambda+
B_H\hat{\overline {H}}^{\alpha}{\hat H}_{\alpha}
+B_{\Sigma}{\hat \Sigma}^{\alpha}_{\beta}
{\hat \Sigma}_{\alpha}^{\beta}
+h_{f}\,\hat{\overline{H}}^{\alpha}
{\hat \Sigma}_{\alpha}^{\beta} {\hat H}_{\beta}\\
& +\frac{h_{\lambda}}{3}\,{\hat \Sigma}_{\alpha}^{\beta}
{\hat \Sigma}_{\beta}^{\gamma}{\hat \Sigma}_{\gamma}^{\alpha}+
\frac{h_{t}}{4}\,
\epsilon^{\alpha\beta\gamma\delta\tau}\,
{\hat \Psi}^{(3)}_{\alpha\beta}
{\hat \Psi}^{(3)}_{\gamma\delta}{\hat H}_{\tau}+
\sqrt{2}h_{b}\,{\hat \Phi}^{(3) \alpha}
{\hat \Psi^{(3)}}_{\alpha\beta}\hat{\overline{H}}^{\beta}
+\mbox{h.c.}\, \}~,
\end{split}
\eeq
where the hat~denotes the scalar components~of the chiral superfields.
The parameters~$M$, $\mu_{\Sigma}$ and $\mu_H$ are treated as independent ones,
since they cannot~be reduced in a suitable form.
The lowest-order~reduction for the parameters of the SSB Lagrangian are~given by:
\beq
\label{red_sol_1}
B_H = \frac{1029}{521}\,\mu_H M~,~
B_{\Sigma}=-\frac{3100}{521}\,\mu_{\Sigma} M~,
\eeq
\beq
\begin{split}
\label{red_sol}
h_t &=-g_t\,M~,~h_b =-g_b\,M~,
~h_f =-g_f\,M~,~h_{\lambda}=0~,\\
m_{H_u}^{2} &=-\frac{569}{521} M^{2}~,~
m_{H_d}^{2} =-\frac{460}{521} M^{2}~,
~m_{\Sigma}^{2} = \frac{1550}{521} M^{2}~,\\
m_{\Phi^3}^{2} & = \frac{436}{521} M^{2}~,~
m_{\Phi^{1,2}}^{2} =\frac{8}{5} M^{2}~,~
m_{\Psi^3}^{2} =\frac{545}{521} M^{2}~,~
m_{\Psi^{1,2}}^{2} =\frac{12}{5} M^{2}~.
\end{split}
\eeq
We choose the gaugino~mass $M$ for characterizing the SUSY breaking scale.
Finally, we note~that (i) $B_{\Sigma}$ and $B_H$ are treated as independent parameters
without spoiling~the one-loop reduction solution of Eq.~(\ref{red_sol}) and
(ii) the soft~scalar mass sum rule still holds despite the specific relations among the gaugino mass and the soft scalar~masses.

We analyze the particle~spectrum predicted
for $\mu < 0$ as the only~phenomenologically acceptable choice (in the $\mu>0$ the quark masses do not match the experimental~measurements).  Below $M_{\rm GUT}$ all couplings and~masses of the theory~run according
to the RGEs of the~MSSM.  Thus we examine~the evolution of these
parameters~according~to their RGEs up to two-loops for~dimensionless 
parameters and at~one-loop~for dimensionful ones imposing the
corresponding~boundary~conditions.

As presented~in \cite{Heinemeyer:2020ftk}, the pole top mass $\mt$ is
predicted within 2$\sigma$ of \refeq{mtmbexp}.  
Concerning the $\mb (M_Z)$ prediction (also in
\cite{Heinemeyer:2020ftk}), we take into account a theoretical 
uncertainty of $\sim 3\%$. But even taking theoretical and experimental
uncertainties into account in combination, we find agreement with
the experimental value only at the 4$\sigma$ level. However,
since there additional uncertainties of a few percent on the quark Yukawa couplings at the SUSY-breaking scale, that were not fully included (see
\cite{Kubo:1995cg})  into the evaluation of  
the bottom mass,
we still consider the model as viable and proceed with its analysis.

\begin{figure}[H]
\centering
\includegraphics[width=0.495\textwidth]{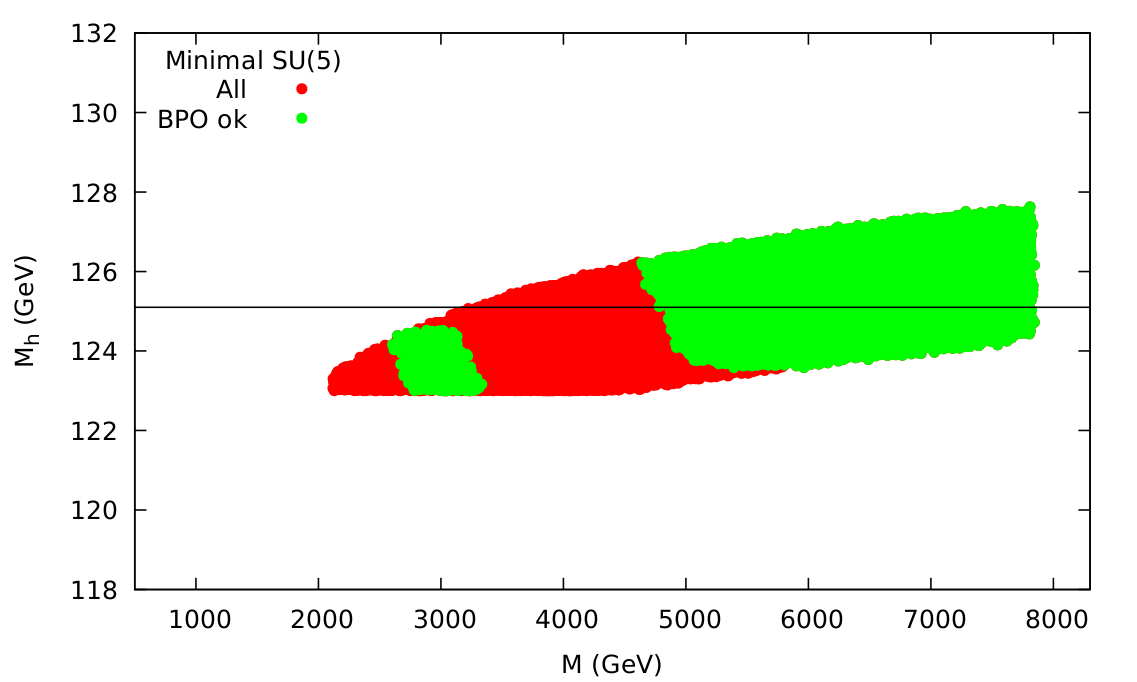}
\includegraphics[width=0.495\textwidth]{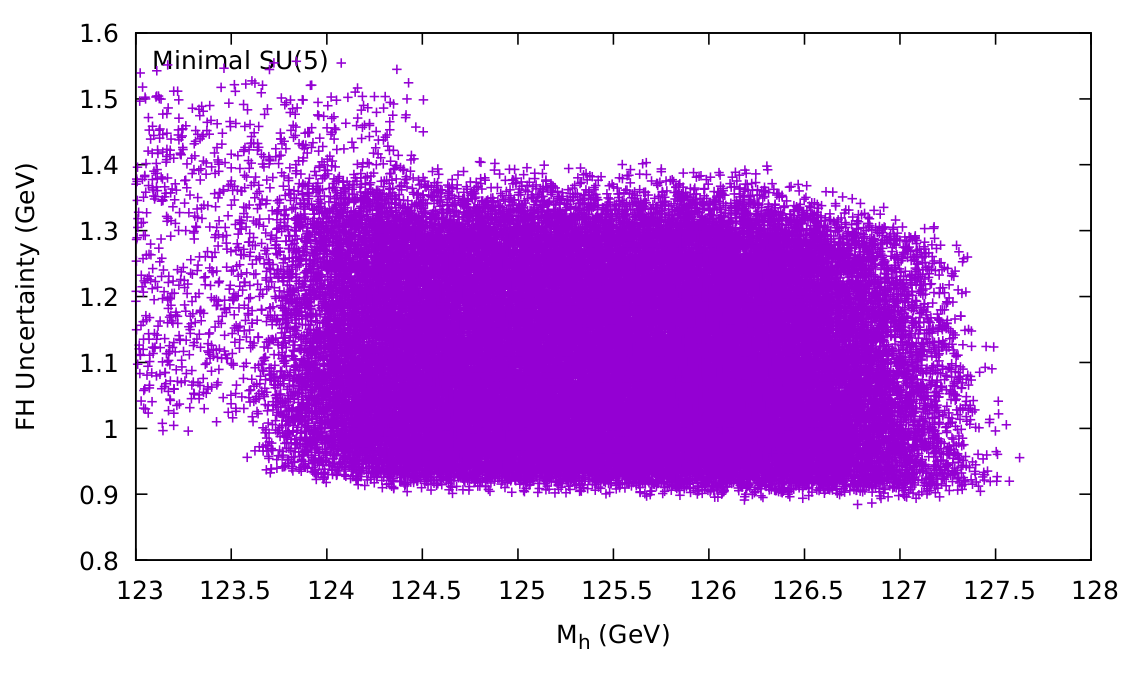}
\caption{Scatter plots for the Minimal $N=1$ $SU(5)$ model. Left: The~lightest Higgs mass, $M_h$, as~a function of $M$.
 The $B$-physics~constraints allow (mostly) higher scale points (with green color).
Right: The theoretical~uncertainty of the light Higgs mass. \cite{Bahl:2019hmm}.}
\label{fig:minhiggsvsM}
\end{figure}

The prediction~for $M_h$ as a function of the unified gaugino mass $M$ with $\mu < 0$ is given in \reffi{fig:minhiggsvsM}  (left).  The $\Delta M_{B_s}$ channel is~responsible for the gap at the $B$-physics allowed points (green points).  
The scattered points come from the fact that  for each $M$ we vary the free parameters $\mu_{\Sigma}$ and $\mu_H$.
 \reffi{fig:minhiggsvsM} (right)~gives the theoretical uncertainty of the Higgs mass for each point, calculated~with {\tt FeynHiggs} 
 2.16.0 \cite{Bahl:2019hmm}. There is substantial improvement to the Higgs mass uncertainty compared~to past analyses, 
 since it has dropped by more~than $1$ GeV.

\small
%%%%%%%%%%%%%%%%%%%% T A B L E %%%%%%%%%%%%%%%%%%%%%%%%%%%%%%%%%%%%%%%%%%%%%%%
\begin{table}[htb!]
\renewcommand{\arraystretch}{1.3}
\centering\small
\begin{tabular}{|c|rrrrrrrrrr|}
\hline
 & $M_1$ & $M_2$ & $M_3$ & $|\mu|$ & $b~~~$ & $A_u$ & $A_d$ & $A_e$ & $\tan{\beta}$ & $m_{Q_{1,2}}^2$  \\
\hline
MINI-1  & 1227 & 2228 & 5310 & 4236 & $401^2$ & 4325 & 4772 & 1732 & 50.3 & $6171^2$  \\
MINI-2  & 1507 & 2721 & 6376 & 5091 & $496^2$  & 5245 & 5586 & 2005 & 52.0 & $7445^2$ \\
MINI-3  & 2249 & 4019 & 9138 & 7367 & $1246^2$  & 7571 & 8317 & 3271 & 50.3 & $10762^2$ \\
\hline
& $m_{Q_{3}}^2$ &  $m_{L_{1,2}}^2$ & $m_{L_{3}}^2$  & $m_{\overline{u}_{1,2}}^2$ & $m_{\overline{u}_{3}}^2$ & $m_{\overline{d}_{1,2}}^2$ & $m_{\overline{d}_{3}}^2$ & $m_{\overline{e}_{1,2}}^2$  & $m_{\overline{e}_{3}}^2$ & \\
\hline
MINI-1 & $4548^2$ & $3714^2$ & $2767^2$ & $5974^2$ & $4181^2$ & $5478^2$ & $4177^2$ & $4160^2$  & $2491^2$   & \\
MINI-2 & $5469^2$ & $4521^2$ & $3358^2$ & $7206^2$ & $5039^2$ & $5478^2$ & $4994^2$ & $5070^2$ & $3019^2$   & \\
MINI-3 & $7890^2$ & $6639^2$ & $4934^2$ & $10412^2$ & $7233^2$ & $9495^2$ & $7211^2$ & $7459^2$ & $4464^2$   & \\
\hline
\end{tabular}
\caption{
Minimal $N=1$ $SU(5)$ predictions that are used as input to {\tt SPheno}.
Mass parameters are in~$\gev$ and rounded to $1 \gev$.}
\label{tab:mininput}
\renewcommand{\arraystretch}{1.0}
\end{table}
\normalsize

Large parts of the predicted particle spectrum
are in agreement~with~the
$B$-physics~observables~and the lightest Higgs boson~mass
measurement~and its theoretical uncertainty. We choose three benchmarks in the low-mass region,
marking the points with the lightest SUSY particle (LSP)~above $1200$~GeV (MINI-1), $1500$~GeV
(MINI-2) and $2200$~GeV (MINI-3), respectively.
The mass of the LSP can go as high as $\sim 3800 \gev$, but the cross sections calculated below will then be negligible and we restrict ourselves here to the low-mass region.
The values~presented in \refta{tab:mininput} were used as input to
get~the full supersymmetric spectrum from {\tt
  SPheno\,4.0.4}~\cite{Porod:2003um,Porod:2011nf}. $M_i$ are the~gaugino
masses and
the rest are~squared soft  sfermion masses which are diagonal
($\mathbf{m^2}=\rm diag(m_1^2,m_2^2,m_3^2)$), and  soft~trilinear
couplings (also diagonal $\mathbf{A_i}=\mathbb{1}_{3\times3}A_i$).

The resulting~masses of all the particles that will be relevant for our
analysis can~be found in \refta{tab:minispheno}. The three first values
are the heavy Higgs masses. The~gluino mass is $M_{\tilde{g}}$, the
neutralinos and the~charginos are denoted as $M_{\tilde{\chi}_i^0}$~and
$M_{\tilde{\chi}_i^{\pm}}$, while the slepton and sneutrino masses
for all three generations are given as
$M_{\tilde{e}_{1,2,3}},~M_{\tilde{\nu}_{1,2,3}}$. Similarly, the squarks
are denoted as $M_{\tilde{d}_{1,2}}$ and $M_{\tilde{u}_{1,2}}$ for the
first two generations. The third generation masses are given by
$M_{\tilde{t}_{1,2}}$ for stops and $M_{\tilde{b}_{1,2}}$ for sbottoms.

%%%%%%%%%%%%%%%%%%%% T A B L E %%%%%%%%%%%%%%%%%%%%%%%%%%%%%%%%%%%%%%%%%%%%%%%
\begin{center}
\begin{table}[ht]
\begin{center}
\small
\begin{tabular}{|l|r|r|r|r|r|r|r|r|r|r|r|}
\hline
 & $M_{H}$ & $M_A$ & $M_{H^{\pm}}$ & $M_{\tilde{g}}$ & $M_{\tilde{\chi}^0_1}$ & $M_{\tilde{\chi}^0_2}$ & $M_{\tilde{\chi}^0_3}$  & $M_{\tilde{\chi}^0_4}$ &  $M_{\tilde{\chi}_1^\pm}$ & $M_{\tilde{\chi}_2^\pm}$ \\\hline
MINI-1 & 2.660 & 2.660 & 2.637 & 5.596 & 1.221 & 2.316 & 4.224 & 4.225 & 2.316 & 4.225  \\\hline
MINI-2 & 3.329 & 3.329 & 3.300 & 6.717 & 1.500 & 2.827 & 5.076 & 5.077 & 2.827 & 5.078  \\\hline
MINI-3 & 8.656 & 8.656 & 8.631 & 9.618 & 2.239 & 4.176 & 7.357 & 7.358 & 4.176 & 7.359  \\\hline
 & $M_{\tilde{e}_{1,2}}$ & $M_{\tilde{\nu}_{1,2}}$ & $M_{\tilde{\tau}}$ & $M_{\tilde{\nu}_{\tau}}$ & $M_{\tilde{d}_{1,2}}$ & $M_{\tilde{u}_{1,2}}$ & $M_{\tilde{b}_{1}}$ & $M_{\tilde{b}_{2}}$ & $M_{\tilde{t}_{1}}$ & $M_{\tilde{t}_{2}}$ \\\hline
MINI-1 & 3.729 & 3.728 & 2.445 & 2.766 & 5.617 & 6.100 & 4.332 & 4.698 & 4.312 & 4.704 \\\hline
MINI-2 & 4.539 & 4.538 & 2.968 & 3.356 & 6.759 & 7.354 & 5.180 & 5.647 & 5.197 & 5.652 \\\hline
MINI-3 & 6.666 & 6.665 & 4.408 & 4.935 & 9.722 & 10.616 & 7.471 & 8.148 & 7.477 & 8.151 \\\hline
\end{tabular}
\caption{Masses of Higgs bosons and some of the SUSY particles for each benchmark of the Minimal $N=1$ $SU(5)$ (in TeV).}\label{tab:minispheno}
\end{center}
\end{table}
\end{center}

At this point there is an important remark. No point fulfills~the
strict~bound of \refeq{cdmexp}, since we have overproduction of CDM in the early universe. The LSP, which in our case is the lightest neutralino, is strongly Bino-like. Combined with the heavy mass it acquires (1-2 TeV), it cannot account for a relic density low enough to agree with experimental observation.
Thus, we need a mechanism that reduces this CDM~abundance. This could~be related~to the~problem~of neutrino~masses, which cannot~be generated~naturally in this particular model. However, one could extend the model by considering bilinear~R-parity
violating terms~and thus introduce~neutrino masses~\cite{Valle:1998bs,Valle3}.
R-parity~violation \cite{herbi}
would have~a small impact~on the~collider phenomenology, but remove~the CDM bound~of
\refeq{cdmexp} completely. ~Other mechanisms, not
involving~R-parity violation, that~could be invoked if the~amount of~CDM appears to~be~too~large, concern~the cosmology~of the early~universe. For~example,
``thermal~inflation'' \cite{thermalinf} or ``late~time entropy~injection'' \cite{latetimeentropy} can~bring the~CDM density into~agreement~with Planck measurements. For the original discussion see \cite{Heinemeyer:2020ftk}.

Table \ref{minSU5xsec} shows the expected production cross section for
selected channels at the  100 TeV future FCC-hh collider. 
We do not show any cross sections for $\sqrt{s} = 14$~TeV, since  the
prospects for discovery of MINI scenarios at the HL-LHC are very
dim. SUSY particles are too heavy to be produced with cross sections
greater that 0.01\,fb.
Concerning the heavy Higgs bosons, the main
search channels will be $H/A \to \tau^+\tau^-$. Our heavy Higgs-boson
mass scale shows values $\gsim 2500$~GeV with $\tb \sim 50$. The
corresponding reach of the HL-LHC has been estimated
in~\cite{HAtautau-HL-LHC}. In comparison with our benchmark points we
conclude that they will not be accessible at the HL-LHC.%
\footnote{The analysis presented in~\cite{HAtautau-HL-LHC} only
  reaches $\MA \le 2000$~GeV, where an exclusion down to
  $\tb \sim 30$ is expected. An extrapolation to $\tb \sim 50$ reaches
  Higgs-boson mass scales of $\sim 2500$~GeV.}

The situation changes for the FCC-hh \cite{fcc-hh}. Theory
analyses~\cite{HAtautau-FCC-hh,Hp-FCC-hh} have shown that for large
$\tb$ heavy Higgs-boson mass scales up to $\sim 8$~TeV may be
accessible, both for neutral as well as for charged Higgs bosons. The
relevant decay channels are $H/A \to \tau^+\tau^-$ and
$H^\pm \to \tau\nu_\tau, tb$. This places our three benchmark points well
within the covered region (MINI-1 and MINI-2) or at the border of the
parameter space that can be probed (MINI-3).

The energy of 100 TeV is big enough to produce SUSY particles in
pairs. However, the cross sections remain relatively small. Only for the
MINI-1 scenario the squark pair and  squark-gluino  (summed over all
squarks) production cross sections can reach tens of fb. For MINI-2 and
MINI-3 scenarios the cross sections are significantly smaller.   In
these scenarios squarks decay preferentially into a quark+LSP (with $\rm
BR\sim 0.95$), gluino into $\tilde t\bar t$ and $\tilde b\bar b$ +$h.c$
with $\rm BR\sim 0.33$ each.   

\begin{center}
\begin{table}[ht]
\begin{center}
\small
\begin{tabular}{|c|c|c|c||c|c|c|c|} \hline
scenarios & MINI-1  & MINI-2 & MINI-3  &  scenarios  &  MINI-1 & MINI-2 & MINI-3\\
$\sqrt{s}$ & 100 TeV & 100 TeV &  100 TeV  &  $\sqrt{s}$  & 100 TeV & 100 TeV &  100 TeV\\ \hline
$\tilde{\chi}^0_1 \tilde{\chi}^0_1 $  &   0.04 & 0.02 & &$\tilde{u}_i \tilde{\chi}^-_1, \tilde{d}_i \tilde{\chi}^+_1 + h.c.$  &   1.00 & 0.35 & 0.03 \\
$\tilde{\chi}^0_1 \tilde{\chi}^0_3 $  &   0.02 & 0.01 & &$\tilde{u}_i \tilde{\chi}^-_2, \tilde{d}_i \tilde{\chi}^+_2 + h.c.$  &   0.07 & 0.02 &     \\
$\tilde{\chi}^0_2 \tilde{\chi}^0_2 $  &   0.06 & 0.02 & &$\tilde{q}_i \tilde{\chi}^0_1, \tilde{q}_i^* \tilde{\chi}^0_1$  &   0.38 & 0.14 & 0.02 \\
$\tilde{\chi}^0_2 \tilde{\chi}^0_3 $  &   0.03 & 0.01 & &$\tilde{q}_i \tilde{\chi}^0_2, \tilde{q}_i^* \tilde{\chi}^0_2$  &   0.51 & 0.17 & 0.02 \\
$\tilde{\chi}^0_2 \tilde{\chi}^0_4 $  &   0.02 & 0.01 & &$\tilde{\nu}_i \tilde{e}_j^*, \tilde{\nu}_i^* \tilde{e}_j$  &   0.06 & 0.02 &     \\
$\tilde{\chi}^0_3 \tilde{\chi}^0_4 $  &   0.05 & 0.02 & & $H  b  \bar{b} $ & 84.04 & 30.10 & 0.17 \\
$\tilde{\chi}^0_2 \tilde{\chi}_1^+ $  &   2.20 & 0.98 & 0.18 & $A b  \bar{b} $ & 84.79 & 29.79 & 0.18\\
$\tilde{\chi}^0_3 \tilde{\chi}_2^+ $  &   0.10 & 0.04 & 0.01 & $H^+ b  \bar{t} + H^- t  \bar{b}$ & 33.24 & 12.76 & 0.1 \\
$\tilde{\chi}^0_4 \tilde{\chi}_2^+ $  &   0.10 & 0.04 & 0.01 & $H^- b  \bar{b} $ & 0.04 & 0.02 & \\
$ \tilde{g}  \tilde{g} $  &   7.76 & 2.02 & 0.11 & $H  t  \bar{t} $ & 0.03 & 0.01 & \\
$ \tilde{g} \tilde{\chi}^0_1 $  &   0.28 & 0.11 & 0.01 & $A t  \bar{t} $ & 0.02 & 0.01 & \\
$ \tilde{g} \tilde{\chi}^0_2 $  &   0.34 & 0.12 & 0.01 & $H  t  b $ & 0.01 &  & \\
$ \tilde{g} \tilde{\chi}_1^+ $  &   0.70 & 0.27 & 0.03 &$ H  A$ & 0.03 & 0.01 &    \\
$\tilde{q}_i \tilde{q}_j, \tilde{q}_i \tilde{q}_j^*$  &   21.15 & 7.44 & 0.74 &$ H  H^+$ & 0.06 & 0.02 &    \\
$\tilde{\chi}_1^+ \tilde{\chi}_1^- $  &   1.19 & 0.54 & 0.09 &$H^+W^-$ & 6.50 & 2.96 & 0.03 \\
$\tilde{\chi}_1^+ \tilde{\chi}_2^- $  &   0.05 & 0.02 & &$ H  W^+$ & 0.02 & 0.01 &    \\
$\tilde{\chi}_2^+ \tilde{\chi}_1^- $  &   0.05 & 0.02 & &$H^+H^-$ & 0.04 & 0.01 &    \\
$\tilde{\chi}_2^+ \tilde{\chi}_2^- $  &   0.06 & 0.02 & &$AH^+$ & 0.06 & 0.02 &    \\
$\tilde{e}_i \tilde{e}_j^*$  &   0.16 & 0.08 & 0.01 &$AW^+$ & 0.02 & 0.01 &    \\
$\tilde{q}_i \tilde{g}, \tilde{q}_i^* \tilde{g}$  &   30.57 & 9.33 & 0.66 &$ H  Z$ & 1.38 & 0.58 & 0.01 \\
$\tilde{\nu}_i \tilde{\nu}_j^*$  &   0.04 & 0.02 & &$AZ$ & 1.20 & 0.52 & 0.01 \\
\hline
\end{tabular}
\caption{Expected production cross sections (in fb) for SUSY particles in the MINI scenarios.  There are no channels with cross sections exceeding  0.01 fb at $\sqrt{s}=14$ TeV.
}
\label{minSU5xsec}
\end{center}
\end{table}
\end{center}

The SUSY discovery reach at the FCC-hh with $3\,\iab$ was evaluated
in~\cite{SUSY-FCC-hh} for a certain set of simplified models. In the
following we will compare these simplified model limits with our
benchmark points to get an idea, which part of the spectrum can be
covered at the FCC-hh. A more detailed evaluation with the future limits
implemented into proper recasting tools would be necessary to obtain a
firmer statement. However, such a detailed analysis goes beyond the
scope of our paper and we restrict ourselves to the simpler direct
comparison of the simplified model limits with our benchmark
predictions.

Concerning the scalar tops, the mass predictions of MINI-1 and MINI-2
are well within the anticipated reach of the FCC-hh, while MINI-3
predicts a too heavy stop mass. On the other hand, even for MINI-1 and
MINI-2 no $5\,\sig$ discovery can be expected.
The situation looks more favorable for the first and second
generation squarks. All the predicted masses can be excluded at the
FCC-hh, whereas a $5\,\sig$ discovery will be difficult, but potentially
possible (see Fig.~19 in \cite{SUSY-FCC-hh}).
Even more favorable appear the prospects for gluino searches at the
FCC-hh. All three benchmark points may lead to a $5\,\sig$ discovery
(see Fig.~13 in \cite{SUSY-FCC-hh}). On the other hand, chances for
chargino/neutralino searches are slim at the FCC-hh. The Next-to LSP
(NLSP) can only be accessed for $\mneu1 \lsim 1$~TeV (see Fig.~21 in
\cite{SUSY-FCC-hh}), where all our benchmark points have
$\mneu1 > 1$~TeV.
Taking into account that our three benchmark points represent only
the lower part of the possible mass spectrum (with LSP masses of up to
$\sim 1.5 \tev$ higher), we conclude that even at the FCC-hh large parts
of the possible SUSY spectrum will remain elusive.

%%%%%%%%%%%%%%%%%%%%%%%%%%%%%%%%%%%%%%%%%%%%%%%%%%%%%%%%%%%%%%%%%%%%%%%%%%%%%%%
%%%%%%%%%%%%%%%%%%%%%%%%%%%%%%%%%%%%%%%%%%%%%%%%%%%%%%%%%%%%%%%%%%%%%%%%%%%%%%%

\section{The Finite $N=1$ Supersymmetric $SU(5)$ Model}\label{sec:finitesu5}
We proceed now to the finite~to all-orders $SU(5)$ gauge theory, where the reduction of couplings
is restricted to the third~generation. An older examination of this specific Finite Unified Theory (FUT) was shown to be in agreement with
the experimental~constraints at the time
\cite{Heinemeyer:2007tz}
and has predicted,~almost five years before its discovery, the light Higgs mass in the correct range. As discussed below,~improved Higgs calculations predict a somewhat different interval that is still in agreement with current~experimental data.
The particle~content of the model has three ($\overline{\bf 5} + \bf{10}$) supermultiplets for the three generations
of leptons~and quarks, while the Higgs sector consists of four supermultiplets
($\overline{\bf 5} + {\bf 5}$) and one ${\bf 24}$. The~finite $SU(5)$ group is broken to the MSSM, which of~course in no longer a~finite theory
\cite{Kapetanakis:1992vx,Kubo:1994bj,Kubo:1994xa,Kubo:1995hm,Kubo:1997fi,Mondragon:1993tw}.

In order for this~finite to all-orders $SU(5)$ model to achieve Gauge Yukawa Unification (GYU),
it should have the~following characteristics:\\
(i) The one-loop~anomalous dimensions are diagonal
i.e., $\gamma_{i}^{(1)\,j} \propto \delta^{j}_{i} $.\\
(ii) The fermions~of the $\overline{\bf 5}_{i}$ and ${\bf 10}_i~(i=1,2,3)$
are not coupled~to the ${\bf 24}$.\\
(iii) The pair~of the MSSM Higgs doublets are mostly composed from the $5$ and $\bar 5$
Higgs that couple~to the third generation

The superpotential of the model,~with an enhanced symmetry due to the reduction of couplings, is given by
\cite{Kobayashi:1997qx,Mondragon:2009zz}:
\begin{align}
W &= \sum_{i=1}^{3}\,[~\frac{1}{2}g_{i}^{u}
\,{\bf 10}_i{\bf 10}_i H_{i}+
g_{i}^{d}\,{\bf 10}_i \overline{\bf 5}_{i}\,
\overline{H}_{i}~] +
g_{23}^{u}\,{\bf 10}_2{\bf 10}_3 H_{4} \\
 &+g_{23}^{d}\,{\bf 10}_2 \overline{\bf 5}_{3}\,
\overline{H}_{4}+
g_{32}^{d}\,{\bf 10}_3 \overline{\bf 5}_{2}\,
\overline{H}_{4}+
g_{2}^{f}\,H_{2}\,
{\bf 24}\,\overline{H}_{2}+ g_{3}^{f}\,H_{3}\,
{\bf 24}\,\overline{H}_{3}+
\frac{g^{\lambda}}{3}\,({\bf 24})^3~.\nonumber
\label{w-futb}
\end{align}

Discussion of~the model with a more detailed description can be found in
\cite{Kapetanakis:1992vx,Kubo:1994bj,Mondragon:1993tw}.
The non-degenerate and~isolated~solutions to the vanishing of $\gamma^{(1)}_{i}$ are:
\begin{equation}
\label{zoup-SOL52}
\begin{split}
& (g_{1}^{u})^2
=\frac{8}{5}~ g^2~, ~(g_{1}^{d})^2
=\frac{6}{5}~g^2~,~
(g_{2}^{u})^2=(g_{3}^{u})^2=\frac{4}{5}~g^2~,\\
& (g_{2}^{d})^2 = (g_{3}^{d})^2=\frac{3}{5}~g^2~,~
(g_{23}^{u})^2 =\frac{4}{5}~g^2~,~
(g_{23}^{d})^2=(g_{32}^{d})^2=\frac{3}{5}~g^2~,\\
& (g^{\lambda})^2 =\frac{15}{7}g^2~,~ (g_{2}^{f})^2
=(g_{3}^{f})^2=\frac{1}{2}~g^2~,~ (g_{1}^{f})^2=0~,~
(g_{4}^{f})^2=0~.
\end{split}
\end{equation}
We have also~the relation $h=-MC$, while the sum rules lead to:
\beq
m^{2}_{H_u}+
2 m^{2}_{{\bf 10}} =M^2~,~
m^{2}_{H_d}-2m^{2}_{{\bf 10}}=-\frac{M^2}{3}~,~%\nonumber\\
m^{2}_{\overline{{\bf 5}}}+
3m^{2}_{{\bf 10}}=\frac{4M^2}{3}~.
\label{sumrB}
\eeq
Therefore, we~only have two free parameters,~namely $m_{{\bf 10}}$ and $M$ in the dimensionful sector.

When $SU(5)$ breaks~down to the MSSM, a suitable rotation in the Higgs sector
\cite{Leon:1985jm,Kapetanakis:1992vx,Mondragon:1993tw,Hamidi:1984gd, Jones:1984qd,Babu:2002in},
permits  only a pair~of Higgs doublets (coupled mostly to the third family) to remain light and acquire vev's.
Avoiding fast proton~decay is achieved with the usual doublet-triplet splitting, although different from
the one applied to~the minimal $SU(5)$ due to the extended Higgs sector of the finite model.
Therefore, below~the GUT scale we get the MSSM where the third generation is given by the finiteness conditions~while the first two remain unrestricted.\\

Conditions~set by finiteness do~not restrict the ~renormalization
properties~at low~energies, so we are left with boundary conditions
on the~gauge~and Yukawa~couplings
(\ref{zoup-SOL52}), the~$h=-MC$
relation~and~the soft
scalar-mass~sum~rule at $M_{\rm GUT}$. The quark masses
$m_b (M_Z)$ and~$m_t$ are predicted within 2$\sigma$ and
  3$\sigma$ uncertainty, respectively of their experimental values
(see \cite{Heinemeyer:2020ftk} for details). 
The only~phenomenologically viable option is to consider $\mu < 0$, as shown in \cite{Heinemeyer:2012yj,Heinemeyer:2012ai,Heinemeyer:2013fga,Heinemeyer:2018zpw,Heinemeyer:2018bab,
Heinemeyer:2019vbc,Heinemeyer:2019nzo,Heinemeyer:2020ftk}.

\begin{figure}[htb!]
\centering
\includegraphics[width=0.43\textwidth]{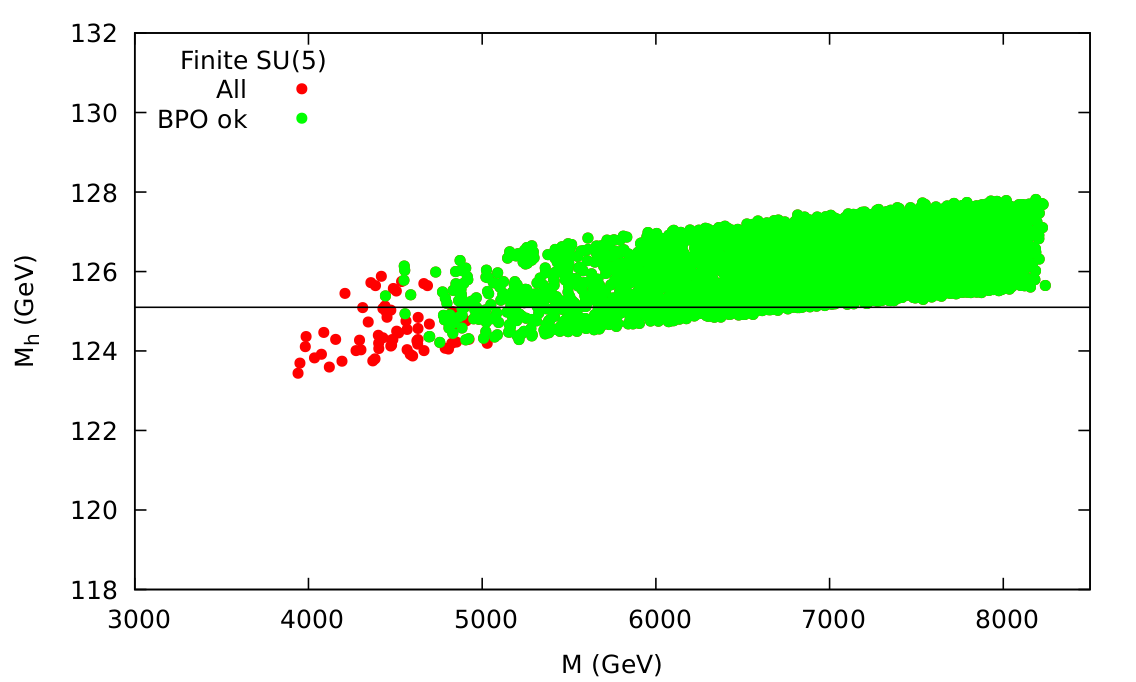}
\includegraphics[width=0.43\textwidth]{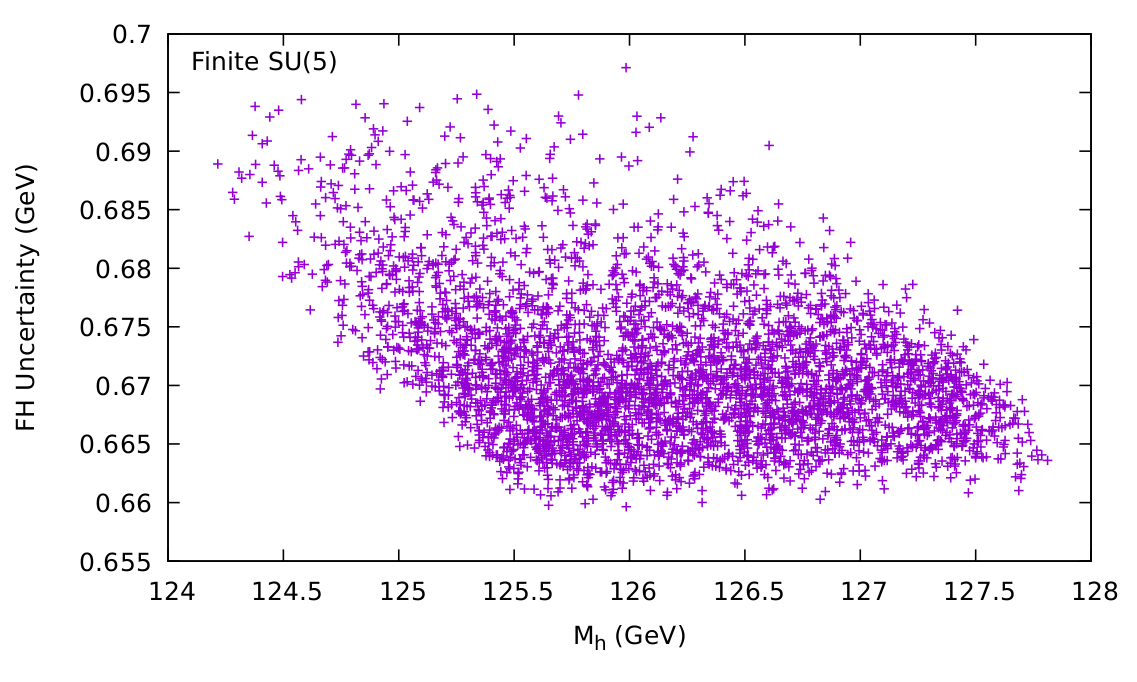}
\caption{Scatter plot~for the Finite $N=1$ $SU(5)$ model. Left: $M_h$ as~a
    function of $M$. Green~points comply with $B$-physics~constraints. Right: The lightest Higgs mass theoretical
    uncertainty calculated~with {\tt FeynHiggs} 2.16.0 \cite{Bahl:2019hmm}.}
\label{fig:futhiggsvsM}
\end{figure}

The scatter plot of the light~Higgs~boson mass is given in \reffi{fig:futhiggsvsM} (left), while~its theory~uncertainty 
\cite{Bahl:2019hmm}~is
given~in \reffi{fig:futhiggsvsM}~(right), with the same color coding as in \reffi{fig:minhiggsvsM}. This point-by-point uncertainty~(calculated with {\tt FeynHiggs})
drops~significantly (w.r.t.~past analyses) to  $0.65-0.70$~GeV. The scattered points come from the free parameter $m_{10}$.

\small
%%%%%%%%%%%%%%%%%%%% T A B L E %%%%%%%%%%%%%%%%%%%%%%%%%%%%%%%%%%%%%%%%%%%%%%%
\begin{table}[htb!]
\renewcommand{\arraystretch}{1.3}
\centering\small
\begin{tabular}{|c|rrrrrrrrrr|}
\hline
& $M_1$ & $M_2$ & $M_3$ & $|\mu|$ &
  $b~~~$ & $A_u$ & $A_d$ & $A_e$ & $\tan{\beta}$ & $m_{Q_{1,2}}^2$ \\
\hline
FUTSU5-1  & 2124 & 3815 & 8804 & 4825 & $854^2$  & 7282 & 7710 & 2961 & 49.9 & $8112^2$ \\
FUTSU5-2  & 2501 & 4473 & 10198 & 5508 & $1048^2$ & 8493 & 9023 & 3536 & 50.1 & $9387^2$ \\
FUTSU5-3  & 3000 & 5340 & 11996 & 6673 & $2361^2$ & 10086 & 10562 & 4243 & 49.9 & $11030^2$ \\
\hline
 & $m_{Q_{3}}^2$ &
  $m_{L_{1,2}}^2$ & $m_{L_{3}}^2$ & $m_{\overline{u}_{1,2}}^2$ & $m_{\overline{u}_{3}}^2$ & $m_{\overline{d}_{1,2}}^2$ & $m_{\overline{d}_{3}}^2$ & $m_{\overline{e}_{1,2}}^2$  & $m_{\overline{e}_{3}}^2$ & \\
\hline
FUTSU5-1  & $6634^2$ & $3869^2$ & $3120^2$ & $7684^2$ & $5053^2$ & $7635^2$ & $4177^2$ & $3084^2$  & $2241^2$ &  \\
FUTSU5-2  & $7669^2$ & $4521^2$ & $3747^2$ & $8887^2$ & $6865^2$ & $8826^2$ & $6893^2$ & $3602^2$  & $2551^2$ &  \\
FUTSU5-3  & $9116^2$ & $5355^2$ & $3745^2$ & $10419^2$ & $8170^2$ & $10362^2$ & $7708^2$ & $4329^2$  & $3403^2$ &  \\
\hline
\end{tabular}
\caption{
Finite $N=1$ $SU(5)$  predictions that are used as input to {\tt SPheno}.
Mass parameters are in~$\gev$ and rounded to $1 \gev$.}
\label{tab:futbinput}
\renewcommand{\arraystretch}{1.0}
\end{table}
\normalsize

Compared~to our previous~analyses
\cite{Heinemeyer:2010xt,Heinemeyer:2012yj,Heinemeyer:2013nza,Heinemeyer:2012ai,Heinemeyer:2013fga,Heinemeyer:2018roq,
Heinemeyer:2018zpw,Heinemeyer:2018bab,Heinemeyer:2019vbc,Heinemeyer:2019nzo,Heinemeyer:2020ftk,Mondragon:2011zzb}, the improved evaluation
of $\Mh$ and its uncertainty~prefer a heavier~(Higgs) spectrum and thus allows only a heavy~supersymmetric spectrum (which is in~agreement with all existing experimental constraints). In particular, very heavy colored SUSY particles are favored (nearly~independent
of the $\Mh$ uncertainty), in~agreement with
the non-observation~of those~particles at the LHC~\cite{2018:59}.

\noindent We choose three~benchmarks, each featuring the LSP above
$2100$~GeV, $2400$~GeV and $2900$~GeV respectively.  Again, they are chosen from the low-mass region.  Although  
 the LSP can be as heavy as $\sim 4000 \gev$, but in such cases the production cross sections even at the FCC-hh would be too small.
The input and output~of {\tt SPheno} 4.0.4
\cite{Porod:2003um,Porod:2011nf} can be found in
\refta{tab:futbinput}~and \refta{tab:futbspheno} (with the notation as
in \refse{sec:minimalsu5}).

Concerning DM, the model~exhibits a high relic abundance for~CDM, as the lightest neutralino (which is the LSP) is again strongly Bino-like (see \cite{Heinemeyer:2020ftk}). The
CDM alternatives proposed~for the Minimal $SU(5)$ model can~also be
applied here. It should be~noted that the bilinear R-parity~violating
terms proposed in the previous~section preserve finiteness,~as well.  

The expected production cross sections for various final states are
listed in Table \ref{futSU5xsec}. 
At 14 TeV HL-LHC none of the  Finite $N=1$ $SU(5)$ scenarios listed in
Table \ref{tab:futbinput} has a SUSY production cross section above 0.01
fb, and thus will (likely) remain unobservable.  All  superpartners are
too heavy to be produced in pairs.  
Also the heavy Higgs bosons are far outside the reach of the
HL-LHC~\cite{HAtautau-HL-LHC}. 

%%%%%%%%%%%%%%%%%%% T A B L E %%%%%%%%%%%%%%%%%%%%%%%%%%%%%%%%%%%%%%%%%%%%%%%
\begin{center}
\begin{table}[ht]
\begin{center}
\small
\begin{tabular}{|l|r|r|r|r|r|r|r|r|r|r|r|r|}
\hline
  & $M_{H}$ & $M_A$ & $M_{H^{\pm}}$  & $M_{\tilde{g}}$ & $M_{\tilde{\chi}^0_1}$ & $M_{\tilde{\chi}^0_2}$ & $M_{\tilde{\chi}^0_3}$  & $M_{\tilde{\chi}^0_4}$ &  $M_{\tilde{\chi}_1^\pm}$ & $M_{\tilde{\chi}_2^\pm}$  \\\hline
FUTSU5-1 & 5.688 & 5.688 & 5.688  & 8.966 & 2.103 & 3.917 & 4.829 & 4.832 & 3.917 & 4.833  \\\hline
FUTSU5-2 & 7.039 & 7.039 &  7.086 & 10.380 & 2.476 & 4.592 & 5.515 & 5.518 & 4.592 & 5.519  \\\hline
FUTSU5-3 & 16.382 & 16.382 &  16.401 & 12.210 & 2.972 & 5.484 & 6.688 & 6.691 & 5.484 & 6.691  \\\hline
 & $M_{\tilde{e}_{1,2}}$ & $M_{\tilde{\nu}_{1,2}}$ & $M_{\tilde{\tau}}$ & $M_{\tilde{\nu}_{\tau}}$ & $M_{\tilde{d}_{1,2}}$ & $M_{\tilde{u}_{1,2}}$ & $M_{\tilde{b}_{1}}$ & $M_{\tilde{b}_{2}}$ & $M_{\tilde{t}_{1}}$ & $M_{\tilde{t}_{2}}$ \\\hline
FUTSU5-1 & 3.102 & 3.907 & 2.205 & 3.137 & 7.839 & 7.888 & 6.102 & 6.817 & 6.099 & 6.821 \\\hline
FUTSU5-2 & 3.623 & 4.566 & 2.517 & 3.768 & 9.059 & 9.119 & 7.113 & 7.877 & 7.032 & 7.881 \\\hline
FUTSU5-3 & 4.334 & 5.418 & 3.426 & 3.834 & 10.635 & 10.699 & 8.000 & 9.387 & 8.401 & 9.390 \\\hline
\end{tabular}
\caption{Masses for each benchmark of the Finite $N=1$ $SU(5)$ (in TeV).}\label{tab:futbspheno}
\end{center}
\end{table}
\end{center}
%%%%%%%%%%%%%%%%%%%%%%%%%%%%%%%%%%%%%%%%%%

At the FCC-hh the discovery prospects for the heavy Higgs-boson
spectrum is significantly better. With $\tb \sim 50$ the first two
benchmark points, FUTSU5-1 and FUTSU5-2, are well within the reach of
the FCC-hh. The third point, FUTSU5-3, however, with $\MA \sim 16$~TeV
will be far outside the reach of the FCC-hh.
Prospects  for  detecting production
of squark pairs and squark-gluino pairs are also very dim since their
production cross section is also at the level of a few fb.  This is as a
result of a heavy spectrum in this class of models
(see \cite{SUSY-FCC-hh} with the same Figures as discussed
in Sec.~\ref{sec:minimalsu5}).
Concerning the stops, the lighter one might be accessible in
FUTSU5-1. For the squarks of the first two generations the prospects
of testing the model are somewhat better. All three benchmark models
could possibly be excluded at the $2\,\sig$ level, but no discovery at
the $5\,\sig$ can be expected. The same holds for the
gluino. Charginos and neutralinos will remain unobservable due to the
heavy LSP. As in the previous section, since only the lower part of
the possible mass spectrum has been considered (with LSP masses higher
by up to $\sim 1 \tev$), we have to conclude that again large parts of
the possible mass spectra will not be observable at the FCC-hh.

\begin{center}
\begin{table}[ht]
\begin{center}
\small
\begin{tabular}{|c|c|c|c||c|c|c|c|}\hline
scenarios &  FUTSU5-1 & FUTSU5-2 & FUTSU5-3  &  scenarios  & FUTSU5-1 & FUTSU5-2 & FUTSU5-3\\
$\sqrt{s}$ &  100 TeV & 100 TeV& 100 TeV  &  $\sqrt{s}$  & 100 TeV & 100 TeV& 100 TeV\\ \hline
$\tilde{\chi}^0_2 \tilde{\chi}^0_3 $ & 0.01 & 0.01 &  & $\tilde{\nu}_i \tilde{\nu}_j^*$ & 0.02 & 0.01 & 0.01 \\
$\tilde{\chi}^0_3 \tilde{\chi}^0_4 $ & 0.03 & 0.01 & & $\tilde{u}_i \tilde{\chi}^-_1, \tilde{d}_i \tilde{\chi}^+_1 + h.c.$ & 0.15 & 0.06 & 0.02 \\
$\tilde{\chi}^0_2 \tilde{\chi}_1^+ $ & 0.17 & 0.08 & 0.03 & $\tilde{q}_i \tilde{\chi}^0_1, \tilde{q}_i^* \tilde{\chi}^0_1$ & 0.08 & 0.03 & 0.01 \\
$\tilde{\chi}^0_3 \tilde{\chi}_2^+ $ & 0.05 & 0.03 & 0.01 & $\tilde{q}_i \tilde{\chi}^0_2, \tilde{q}_i^* \tilde{\chi}^0_2$ & 0.08 & 0.03 & 0.01 \\
$\tilde{\chi}^0_4 \tilde{\chi}_2^+ $ & 0.05 & 0.03 & 0.01 & $\tilde{\nu}_i \tilde{e}_j^*, \tilde{\nu}_i^* \tilde{e}_j$ & 0.09 & 0.04 & 0.01 \\
$ \tilde{g}  \tilde{g} $ & 0.20 & 0.05 & 0.01 & $H  b  \bar{b} $ & 2.76 & 0.85 &  \\
$ \tilde{g} \tilde{\chi}^0_1 $ & 0.03 & 0.01 & & $A b  \bar{b} $ & 2.73 & 0.84 &  \\
$ \tilde{g} \tilde{\chi}^0_2 $ & 0.03 & 0.01 & & $H^+ b  \bar{t} + h.c.$ & 1.32 & 0.42 & \\
$ \tilde{g} \tilde{\chi}_1^+ $ & 0.07 & 0.03 & 0.01 & $H^+W^-$ & 0.38 & 0.12 &    \\
$\tilde{q}_i \tilde{q}_j, \tilde{q}_i \tilde{q}_j^*$ & 3.70 & 1.51 & 0.53 & $H Z$ & 0.09 & 0.03 &    \\
$\tilde{\chi}_1^+ \tilde{\chi}_1^- $ & 0.10 & 0.05 & 0.02 & $AZ$ & 0.09 & 0.03 &    \\
$\tilde{\chi}_2^+ \tilde{\chi}_2^- $ & 0.03 & 0.02 & 0.01 & & & &\\
$\tilde{e}_i \tilde{e}_j^*$ & 0.23 & 0.13 & 0.05 & & & &\\
$\tilde{q}_i \tilde{g}, \tilde{q}_i^* \tilde{g}$ & 2.26 & 0.75 & 0.20 &     &  &  &   \\
 \hline
\end{tabular}
\caption{Expected production cross sections (in fb) for SUSY particles
  in the FUTSU5 scenarios.
}
\label{futSU5xsec}
\end{center}
\end{table}
\end{center}

%%%%%%%%%%%%%%%%%%%%%%%%%%%%%%%%%%%%%%%%%%%%%%%%%%%%%%%%%%%%%%%%%%%%%%%%%%%%%%%
%%%%%%%%%%%%%%%%%%%%%%%%%%%%%%%%%%%%%%%%%%%%%%%%%%%%%%%%%%%%%%%%%%%%%%%%%%%%%%%

\section{The Finite $SU(N)^3$ Model}\label{sec:su33}
We proceed now to  a FUT~based on a product gauge group.
Consider an $N=1$ SUSY theory~with $SU(N)_1 \times SU(N)_2 \times \cdots \times SU(N)_k$
having $n_f$ families~transforming as $(N,N^*,1,\dots,1) + (1,N,N^*,\dots,1) + \cdots + (N^*,1,1,\dots,N)$.
Then, the first order~coefficient of the $\beta$-function, for each $SU(N)$ group is:
\begin{equation}
b = \left( -\frac{11}{3} + \frac{2}{3} \right) N + n_f \left( \frac{2}{3}
 + \frac{1}{3} \right) \left( \frac{1}{2} \right) 2 N = -3 N + n_f
N\,.
\label{3gen}
\end{equation}
Demanding the vanishing of the gauge one-loop $\beta$-function,~i.e. $b=0$, we are led to the choice $n_f = 3$.
Phenomenological reasons~lead to the choice of the $SU(3)_C \times SU(3)_L \times SU(3)_R$ model,
discussed~in Ref.\cite{Ma:2004mi},
while a detailed~discussion of the general well known example can be found in
\cite{Derujula:1984gu,Lazarides:1993sn,Lazarides:1993uw,Ma:1986we}.
The leptons and~quarks transform as:
\begin{equation}
  q = \begin{pmatrix} d & u & D \\ d & u & D \\ d & u & D \end{pmatrix}
\sim (3,3^*,1), ~~~~~~
    q^c = \begin{pmatrix} d^c & d^c & d^c \\ u^c & u^c & u^c \\ D^c & D^c & D^c
\end{pmatrix}
    \sim (3^*,1,3), ~~~~~~ \lambda = \begin{pmatrix} N & E^c & \nu \\ E & N^c & e \\ \nu^c & e^c & S
\end{pmatrix}
\sim (1,3,3^*)
\label{2quarks}
\end{equation}
where $D$ are down-type~quarks acquiring masses~close to $M_{\rm GUT}$.
A cyclic $Z_3$ symmetry~is imposed on the multiplets to achieve equal gauge couplings
at the GUT scale and in~that case the vanishing of the first-order $\beta$-function is satisfied.
Continuing to the~vanishing of the anomalous dimension of all the fields (see Eq.~(\ref{2nd})),
we note that there~are two trilinear invariant terms in the superpotential, namely:
\begin{equation}
f ~Tr (\lambda q^c q) + \frac{1}{6} f' ~\epsilon_{ijk} \epsilon_{abc}
(\lambda_{ia} \lambda_{jb} \lambda_{kc} + q^c_{ia} q^c_{jb} q^c_{kc} +
q_{ia} q_{jb} q_{kc}),
\label{16}
\end{equation}
with $f$ and $f'$~the corresponding Yukawa couplings.
The~superfields $(\tilde N,\tilde N^c)$ obtain vev's and
provide masses to~leptons and quarks
\begin{equation}
m_d = f \langle \tilde N \rangle, ~~ m_u = f \langle \tilde N^c \rangle, ~~
m_e = f' \langle \tilde N \rangle, ~~ m_\nu = f' \langle \tilde N^c \rangle.
\label{18}
\end{equation}
Having three~families, 11 $f$ couplings and 10 $f'$ couplings are present in the most
general superpotential.~Demanding the vanishing of all superfield anomalous dimensions,
9~conditions~are~imposed
\begin{equation}
\sum_{j,k} f_{ijk} (f_{ljk})^* + \frac{2}{3} \sum_{j,k} f'_{ijk}
(f'_{ljk})^* = \frac{16}{9} g^2 \delta_{il}\,,
\label{19}
\end{equation}
where
\begin{equation}
 f_{ijk} = f_{jki} = f_{kij}, ~~~~~~~~  f'_{ijk} = f'_{jki} = f'_{kij} = f'_{ikj} = f'_{kji} = f'_{jik}~.
\end{equation}
The masses of leptons~and quarks are acquired from the vev's of the scalar parts of the superfields $\tilde N_{1,2,3}$~and~$\tilde N^c_{1,2,3}$.

At $M_{\rm GUT}$~the $SU(3)^3$ FUT breaks\footnote{\cite{Irges:2011de,Irges:2012ze} and refs~therein discuss in detail~the~spontaneous breaking of $SU(3)^3$.} to the MSSM,
where as was~already mentioned, both Higgs doublets couple mostly to the third generation.
The FUT breaking~leaves its mark in the form of Eq.~(\ref{19}), i.e. boundary conditions on the gauge and Yukawa couplings,
the~relation among the soft trilinear coupling, the corresponding Yukawa coupling and the unified gaugino mass and finally the soft scalar mass sum rule at $M_{\rm GUT}$. In this specific model
the sum rule~takes the form:
\begin{equation}
m^2_{H_u} + m^2_{\tilde t^c} + m^2_{\tilde q} = M^2 =
m^2_{H_d} + m^2_{\tilde b^c} + m^2_{\tilde q}~.
\end{equation}

The model is finite~to all-orders if the solution of Eq.~(\ref{19}) is both \textit{isolated} and unique.
Then, $f'=0$  and we~have the relations
\begin{equation}
f^2 = f^2_{111} = f^2_{222} = f^2_{333} = \frac{16}{9} g^2\,.
\label{isosol}
\end{equation}
Since all $f'$ vanish,~at one-loop order, the lepton masses vanish. Since these masses,  even radiatively,
cannot be produced~because of the finiteness conditions, we are faced with a problem which needs further study.
If the solution~of Eq.~(\ref{19}) is unique but not isolated (i.e. parametric), we can have non zero $f'$
leading to~non-vanishing lepton masses and at the same time achieving two-loop finiteness.
In that~case the set of conditions restricting the~Yukawa couplings read:
\begin{equation}
f^2 = r \left(\frac{16}{9}\right) g^2\,,\quad
f'^2 = (1-r) \left(\frac{8}{3}\right) g^2\,,
\label{fprime}
\end{equation}
where $r$ parametrises~the different solutions and as such is a free parameter.
It should be noted that~we use the sum rule as boundary condition for the soft scalar masses.\\

In our analysis we~consider the two-loop finite version of the model, where again below $M_{\rm GUT}$ we get the MSSM.
We take into account~two new thresholds for the masses of the new particles at $\sim10^{13}$~GeV and $\sim 10^{14}$~GeV resulting~in a wider phenomenologically viable parameter
space~\cite{Mondragon:2011zzb}.

Looking for the~values of the~parameter
$r$  which comply~with the~experimental limits, we find  that both the top and bottom masses are in~the experimental range~(within 2$\sigma$) for the same
value of $r$ between~$0.65$ and $0.80$ (we singled out the $\mu < 0$ case as the most promising).
The inclusion of the~above-mentioned thresholds gives an important improvement on the top mass from past
versions of the model \cite{Ma:2004mi,Heinemeyer:2009zs,Heinemeyer:2010zzb,Heinemeyer:2010zza}.

\begin{figure}[htb!]
\centering
\includegraphics[width=0.43\textwidth]{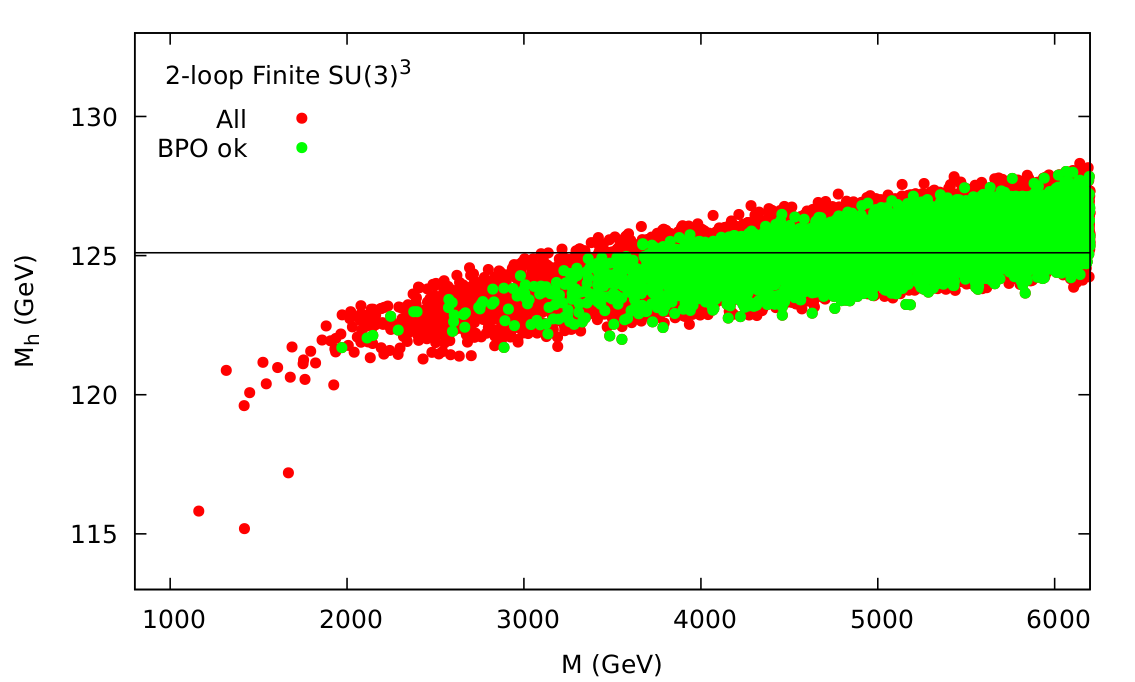}
\includegraphics[width=0.43\textwidth]{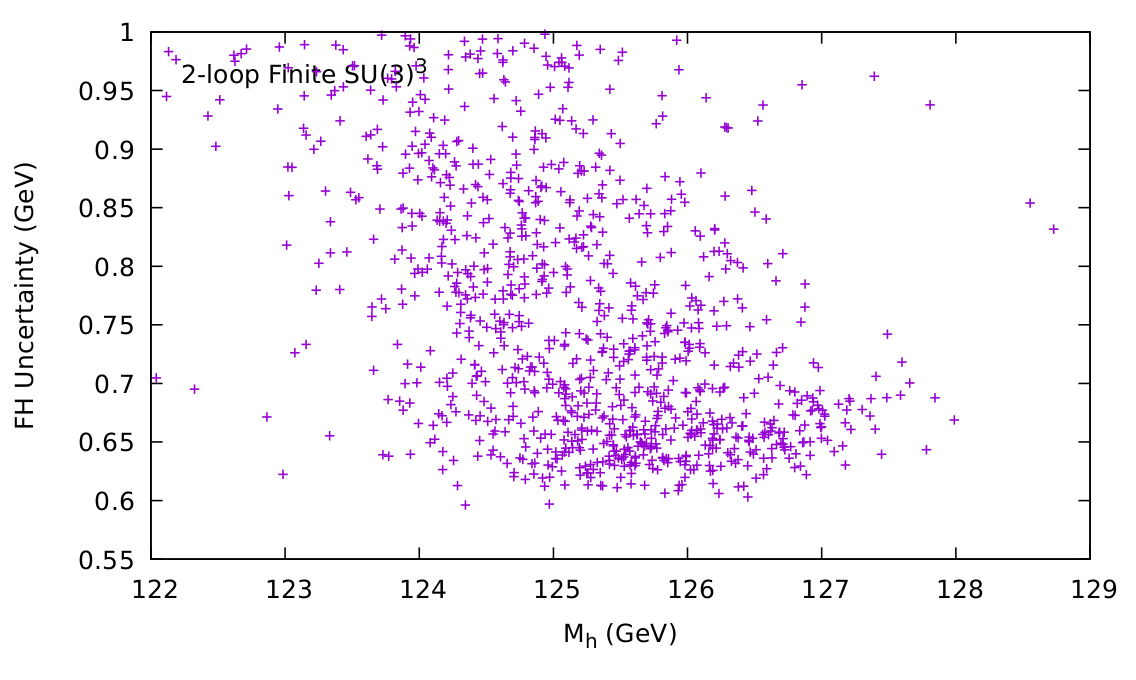}
\caption{Scatter plot for the Finite $N=1$ $SU(3)^3$ model. Left: $M_h$ as~a
    function of $M$. Right: The~Higgs mass theoretical~uncertainty \cite{Bahl:2019hmm}.}
\label{fig:su33higgsvsM}
\end{figure}

 \reffi{fig:su33higgsvsM} (left)  shows the scatter plot of~the light Higgs boson mass (green points satisfy the 
B-physics constraints),
while the point-by-point~calculated theoretical uncertainty is presented in
\reffi{fig:su33higgsvsM}~(right). The scattered points are due to the fact that we vary five parameters, namely $r$ and four of the parameters that form the sum rule. The uncertainty is found in the range between 0.6~GeV and 1.0~GeV.
All constraints regarding~quark masses, the light Higgs boson mass and
B-physics are satisfied, rendering  the model very successful. The
prediction of the SUSY spectrum results in relatively heavy particles,
in full agreement with the current experimental searches. 

\small
%%%%%%%%%%%%%%%%%%%% T A B L E %%%%%%%%%%%%%%%%%%%%%%%%%%%%%%%%%%%%%%%%%%%%%%%
\begin{table}[htb!]
\renewcommand{\arraystretch}{1.3}
\centering\small
\begin{tabular}{|c|rrrrrrrrrr|}
\hline
  & $M_1$ & $M_2$ & $M_3$ & $|\mu|$ &
  $b~~~$ & $A_u$ & $A_d$ & $A_e$ & 
$\tan{\beta}$ & $m_{Q_{1,2}}^2$ \\
\hline
FSU33-1  & 1522 & 2758 & 6369 & 6138 & $1002^2$ & 4520 & 4413 & 1645 & 46.2 & $5574^2$  \\
FSU33-2  & 2070 & 3722 & 8330 & 7129 & $1083^2$ & 5841 & 5734 & 2357 & 45.5 & $7255^2$  \\
FSU33-3  & 2500 & 4484 & 10016 & 6790 & $972^2$ & 7205 & 7110 & 2674 & 49.7 & $8709$  \\
\hline
 & $m_{Q_{3}}^2$ &
  $m_{L_{1,2}}^2$ & $m_{L_{3}}^2$ & $m_{\overline{u}_{1,2}}^2$ & $m_{\overline{u}_{3}}^2$ & $m_{\overline{d}_{1,2}}^2$ & $m_{\overline{d}_{3}}^2$ & $m_{\overline{e}_{1,2}}^2$  & $m_{\overline{e}_{3}}^2$ & \\
\hline
FSU33-1  & $4705^2$ & $2382^2$ & $3754^2$ & $5234^2$ & $5548^2$ & $5197^2$ & $7043^2$ & $1558^2$  & $3095^2$ &  \\
FSU33-2  & $7255^2$ & $3136^2$ & $4131^2$ & $6749^2$ & $7225^2$ & $6745^2$ & $8523^2$ & $2238^2$  & $3342^2$ &  \\
FSU33-3  & $9074^2$ & $3831^2$ & $5483^2$ & $8152^2$ & $7207^2$ & $2558^2$ & $8600^2$ & $2507^2$  & $4000^2$ &  \\
\hline
\end{tabular}
\caption{
Finite $N=1$ $SU(3)^3$ predictions that are used as input to {\tt SPheno}.
Mass parameters are in~$\gev$  and rounded to $1 \gev$.}
\label{tab:su33input}
\renewcommand{\arraystretch}{1.0}
\end{table}
\normalsize

Again, we choose three~benchmarks, each featuring the  LSP above
$1500$~GeV, $2000$~GeV and $2400$~GeV~respectively (but the LSP can go as high as $\sim 4100 \gev$, again with too small cross sections).
The input~and output of {\tt SPheno} 4.0.4
\cite{Porod:2003um,Porod:2011nf} can be found in
\refta{tab:su33input}~and \refta{tab:su33spheno}  respectively
(with the notation as in \refse{sec:minimalsu5}).

Like the previous models, the CDM relic density fails to comply with~the
experimental~bounds (see \refeq{cdmexp}). The
lightest neutralino is the LSP and considered as a~CDM candidate, but
its relic density does not go below $0.15$, since~it is strongly
Bino-like and would require a~lower scale of the particle spectrum. It
should be noted that if the B-physics~constraints allowed for a unified
gaugino mass $\sim 0.5 \tev$ lower,~then agreement with the CDM bounds
as well could be achieved (see \cite{Heinemeyer:2020ftk}). However, the alternatives proposed in the previous sections can be applied in this case as well, preserving finiteness.

%%%%%%%%%%%%%%%%%%%% T A B L E %%%%%%%%%%%%%%%%%%%%%%%%%%%%%%%%%%%%%%%%%%%%%%%
\begin{center}
\begin{table}[ht]
\begin{center}
\small
\begin{tabular}{|l|r|r|r|r|r|r|r|r|r|r|}
\hline
   & $M_{H}$ & $M_A$ & $M_{H^{\pm}}$ & $M_{\tilde{g}}$ & $M_{\tilde{\chi}^0_1}$ & $M_{\tilde{\chi}^0_2}$ & $M_{\tilde{\chi}^0_3}$  & $M_{\tilde{\chi}^0_4}$ &  $M_{\tilde{\chi}_1^\pm}$ & $M_{\tilde{\chi}_2^\pm}$ \\\hline
FSU33-1 & 7.029 & 7.029 & 7.028  & 6.526 & 1.506 & 2.840 & 6.108 & 6.109 & 2.839 & 6.109  \\\hline
FSU33-2 & 6.484 & 6.484 & 6.431  & 8.561 & 2.041 & 3.817 & 7.092 & 7.093 & 3.817 & 7.093  \\\hline
FSU33-3 & 6.539 & 6.539 & 6.590  & 10.159 & 2.473 & 4.598 & 6.780 & 6.781 & 4.598 & 6.781  \\\hline
 & $M_{\tilde{e}_{1,2}}$ & $M_{\tilde{\nu}_{1,2}}$ & $M_{\tilde{\tau}}$ & $M_{\tilde{\nu}_{\tau}}$ & $M_{\tilde{d}_{1,2}}$ & $M_{\tilde{u}_{1,2}}$ & $M_{\tilde{b}_{1}}$ & $M_{\tilde{b}_{2}}$ & $M_{\tilde{t}_{1}}$ & $M_{\tilde{t}_{2}}$ \\\hline
FSU33-1 & 2.416 & 2.415 & 1.578 & 2.414 & 5.375 & 5.411 & 4.913 & 5.375 & 4.912 & 5.411 \\\hline
FSU33-2 & 3.188 & 3.187 & 2.269 & 3.186 & 7.026 & 7.029 & 6.006 & 7.026 & 6.005 & 7.029 \\\hline
FSU33-3 & 3.883 & 3.882 & 2.540 & 3.882 & 8.334 & 8.397 & 7.227 & 8.334 & 7.214 & 7.409 \\\hline
\end{tabular}
\caption{Masses for each benchmark of the Finite $N=1$ $SU(3)^3$ (in TeV).}\label{tab:su33spheno}
\end{center}
\end{table}
\end{center}
 %%%%%%%%%%%%%%%%%%%%%%%%%%%%%%%%%%%%%%%%%%

It should be noted that in this model the scale of the heavy Higgs
bosons does not vary monotonously with $\mneu1$, as in the previously
considered models. This can be understood as follows.
 The Higgs bosons masses are determined by a combination of the
sum rule at 
the unification scale, and the requirement of successful electroweak
symmetry breaking at the low scale.
Like in the finite scenario of
the previous section, there are no direct relations between the soft
scalar masses and the unified gaugino mass,  but they are related through the corresponding sum rule and thus vary correlatedly, a fact that makes the dependence on
the boundary values more restrictive. Furthermore (and even more
importantly), the fact that we took into account the two thresholds
at $\sim 10^{13} \gev$ and $\sim 10^{14} \gev$ (as mentioned above), 
allows the new particles, mainly the
Higgsinos of the two other families (that were considered decoupled at
the unification scale in previous analyses) and the down-like exotic
quarks (in a lower degree), to affect the running of the (soft) RGEs
in a non-negligible way. 
Thus, since at low energies the heavy
Higgs masses depend mainly on 
the values of $m^2_{H_u}$, $m^2_{H_d}$, $|\mu|$ and $\tb$,
they are substantially less connected to $\mneu1$ than
  in the other models, leading to a different exclusion potential, as
  will be discussed in the following.

Scenarios of Finite $SU(3)^3$  are beyond the reach of the HL-LHC. Not
only superpartners are too heavy, but also  heavy Higgs bosons 
with a mass scale of $\sim 7$~TeV cannot be detected at the HL-LHC.
At 100 TeV collider (see \refta{fSU33xsec}),
on the other hand, all three benchmark points are well within the
reach of the $H/A \to \tau^+\tau^-$ as well as the
$H^\pm \to \tau\nu_\tau, tb$ searches~\cite{HAtautau-FCC-hh,Hp-FCC-hh},
despite the slightly smaller values of $\tb \sim 45$. This is
particularly because of the different dependence of the heavy
Higgs-boson mass scale on $\mneu1$, as discussed above.
However, we have checked that $\MA$ can go up to to $\sim 11 \tev$,
and thus the heaviest part of the possible spectrum would escape the
heavy Higgs-boson searches at the FCC-hh.

Interesting are also the prospects for production of squark
pairs and squark-gluino, which 
can reach $\sim 20$ fb for the FSU33-1 case, going down to a few fb for
FSU33-2 and FSU33-3 scenarios.  The lightest squarks decay almost
exclusively  to the third generation quark and chargino/neutralino,
while gluino enjoys many possible decay channels to quark-squark pairs
each one with branching fraction of the order of a percent, with the
biggest one $\sim 20\%$  to $t\tilde{t}_1 +h.c.$.

We briefly discuss the SUSY discovery potential at the FCC-hh,
referring agian to \cite{SUSY-FCC-hh} with the same Figures as discussed
in Sec.~\ref{sec:minimalsu5}.
Stops in FSU33-1 and FSU33-2 can be tested at the FCC-hh, while the
masses turn out to be too heavy in FSU33-3. The situation is better
for scalar quarks, where all three scenarios can be tested, but will
not allow for a $5\,\sig$ discovery. Even more favorable are the
prospects for gluino. Possibly all three scenarios can be tested at
the $5\,\sig$ level. As in the previous scenario, the charginos and
neutralinos will not be accessible, due to the too heavy LSP.
Keeping in mind that only the lower part of possible mass spectrum
  is represented by the three benchmarks (with the LSP up to $\sim 1.5 \tev$
heavier), we conclude that as before large parts of the parameter space
will not be testable at the FCC-hh. The only partial exception here is the
Higgs-boson sector, where only the the part with the highest possible
Higgs-boson mass spectra would escape the FCC-hh searches.

\begin{center}
\begin{table}[ht]
\begin{center}
\small
\begin{tabular}{|c|c|c|c||c|c|c|c|}\hline
scenarios  & FSU33-1 & FSU33-2 & FSU33-3  &  scenarios &  FSU33-1 & FSU33-2 & FSU33-3\\
$\sqrt{s}$  & 100 TeV & 100 TeV& 100 TeV  &  $\sqrt{s}$ &  100 TeV & 100 TeV& 100 TeV\\ \hline
$\tilde{\chi}^0_1 \tilde{\chi}^0_1 $  & 0.04 & 0.01 & 0.01 & $\tilde{q}_i \tilde{g}, \tilde{q}_i^* \tilde{g}$  & 22.12 & 3.71 & 1.05\\
$\tilde{\chi}^0_2 \tilde{\chi}^0_2 $  & 0.04 & 0.01 &  & $\tilde{\nu}_i \tilde{\nu}_j^*$  & 0.10 & 0.03 & 0.01\\
$\tilde{\chi}^0_2 \tilde{\chi}_1^+ $  & 0.58 & 0.16 & 0.07 & $\tilde{u}_i \tilde{\chi}^-_1, \tilde{d}_i \tilde{\chi}^+_1 + h.c.$  & 1.22 & 0.25 & 0.08\\
$\tilde{\chi}^0_3 \tilde{\chi}_2^+ $  & 0.02 & 0.01 & 0.01 & $\tilde{q}_i \tilde{\chi}^0_1, \tilde{q}_i^* \tilde{\chi}^0_1$  & 0.55 & 0.13 & 0.05\\
$\tilde{\chi}^0_4 \tilde{\chi}_2^+ $  & 0.02 & 0.01 & 0.01 & $\tilde{q}_i \tilde{\chi}^0_2, \tilde{q}_i^* \tilde{\chi}^0_2$  & 0.60 & 0.13 & 0.04\\
$ \tilde{g}  \tilde{g} $  & 2.61 & 0.30 & 0.07 & $\tilde{\nu}_i \tilde{e}_j^*, \tilde{\nu}_i^* \tilde{e}_j$  & 0.36 & 0.12 & 0.04\\
$ \tilde{g} \tilde{\chi}^0_1 $  & 0.20 & 0.05 & 0.02 & $H  b  \bar{b} $ & 0.71 & 1.23 & 1.19 \\
$ \tilde{g} \tilde{\chi}^0_2 $  & 0.20 & 0.04 & 0.01 & $A b  \bar{b} $ & 0.72 & 1.23 & 1.18 \\
$ \tilde{g} \tilde{\chi}_1^+ $  & 0.42 & 0.09 & 0.03 & $H^+ b  \bar{t} + h.c.$ & 0.37 & 0.75 & 0.58 \\
$\tilde{q}_i \tilde{q}_j, \tilde{q}_i \tilde{q}_j^*$  & 25.09 & 6.09 & 2.25 & $H^+W^{-}$ & 0.10 & 0.25 & 0.19 \\
$\tilde{\chi}_1^+ \tilde{\chi}_1^- $  & 0.37 & 0.10 & 0.04 & $H Z$ & 0.02 & 0.04 & 0.04 \\
$\tilde{e}_i \tilde{e}_j^*$  & 0.39 & 0.12 & 0.06 & $AZ$ & 0.02 & 0.04 & 0.04 \\
\hline
\end{tabular}
\caption{Expected production cross sections (in fb) for SUSY particles in the FSU33 scenarios.
}
\label{fSU33xsec}
\end{center}
\end{table}
\end{center}

%%%%%%%%%%%%%%%%%%%%%%%%%%%%%%%%%%%%%%%%%%%%%%%%%%%%%%%%%%%%%%%%%%%%%%%%%%%%%%%
%%%%%%%%%%%%%%%%%%%%%%%%%%%%%%%%%%%%%%%%%%%%%%%%%%%%%%%%%%%%%%%%%%%%%%%%%%%%%%%

\section{The Reduced MSSM }\label{sec:mssm}
We finish our phenomenological~analyses with the application of the method of coupling reduction to a version of the MSSM, where a~covering GUT is assumed.
The original partial reduction~can be found in refs.\cite{Mondragon:2013aea,Mondragon:2017hki} where only the third
fermionic~generation is considered. Following this restriction, the superpotential reads:
\beq
\label{supot2}
W = Y_tH_2Qt^c+Y_bH_1Qb^c+Y_\tau H_1L\tau^c+ \mu H_1H_2\, ,
\eeq
where $Y_{t,b,\tau}$ refer only to the third family, and the SSB
Lagrangian is given by by (with the trilinear couplings $h_{t,b,\tau}$
for the third family) 
\beq
\label{SSB_L}
\begin{split}
-\mathcal{L}_{\rm SSB} &= \sum_\phi m^2_\phi\hat{\phi^*}\hat{\phi}+
\left[m^2_3\hat{H_1}\hat{H_2}+\sum_{i=1}^3 \frac 12 M_i\lambda_i\lambda_i +\textrm{h.c}\right]\\
&+\left[h_t\hat{H_2}\hat{Q}\hat{t^c}+h_b\hat{H_1}\hat{Q}\hat{b^c}+h_\tau \hat{H_1}\hat{L}\hat{\tau^c}+\textrm{h.c.}\right] .
%%%%%\qquad       (45)
\end{split}
\eeq

We start~with the dimensionless sector and consider initially the top and bottom Yuakwa couplings and the strong~gauge coupling.
The rest~of the couplings will be treated as corrections.
If~$Y_{(t,b)}^2/(4\pi)\equiv \alpha_{(t,b)}$, the REs and the Yukawa RGEs give
\[
\al_i=G_i^2\al_3, \text{~~~where~~~}  G_i^2=\frac 13       ,\qquad i=t,b.
\]
If the tau~Yukawa is included in the reduction, the corresponding $G^2$ coefficient for tau turns negative \cite{{MTZ:14}},
explaining~why this coupling is treated also as a correction (i.e. it cannot be reduced).

We assume~that the ratios of the top and bottom Yukawa to the strong coupling are constant at the GUT scale, i.e. they have~negligible scale dependence,
\[
\frac{d}{dg_3}\left(\frac {Y_{t,b}^2}{g_3^2}\right)=0.
\]
Then, including the~corrections from the $SU(2)$, $U(1)$ and tau couplings, at the GUT scale, the
coefficients~$G^2_{t,b}$ become:
\beq
\label{Gt2_Gb2}
G_t^2=\frac 13+\frac{71}{525}\rho_1+\frac 37 \rho_2 +\frac 1{35}\rho_\tau,\qquad
G_b^2=\frac 13+\frac{29}{525}\rho_1+\frac 37 \rho_2 -\frac 6{35}\rho_\tau~,
\eeq
where
\beq
\label{r1_r2_rtau}
\rho_{1,2}=\frac{g_{1,2}^2}{g_3^2}=\frac{\al_{1,2}}{\al_3},\qquad
\rho_\tau=\frac{g_\tau^2}{g_3^2}=\frac{\displaystyle{\frac{Y^2_\tau}{4\pi}}}{\al_3}~.
\eeq
We shall treat~Eqs.(\ref{Gt2_Gb2}) as boundary conditions at the GUT scale.

Going to the~two-loop level, we assume that the corrections take the following form:
\[
\al_i=G_i^2\al_3+J_i^2 \al_3^2,\qquad i=t,b~.
\]
Then, the~two-loop coefficients, $J_i$, including the corrections from the gauge and the tau Yukawa couplings,~are:
\[
J_t^2=\frac 1{4\pi}\frac{N_t}{D},\quad
J_b^2=\frac 1{4\pi}\frac{N_b}{5D},
\]
where $D$, $N_t$ and $N_b$ are known~quantities which can be found in ref.\cite{Heinemeyer:2017gsv}.

Proceeding to the the SSB~Lagrangian, Eq.~(\ref{SSB_L}), and the dimension-one parameters, i.e the trilinear couplings~$h_{t,b,\tau}$,
we first~reduce $h_{t,b}$ and we get
\[
h_i=c_i Y_i M_3 = c_i G_i M_3 g_3, \text{~~~where~~~} c_i=-1   \qquad i=t,b,
\]
where $M_3$ is the~gluino~mass. Adding the corrections from the gauge and the tau~couplings we have
\[
c_t=-\frac{A_A A_{bb} + A_{tb} B_B}{A_{bt} A_{tb} - A_{bb} A_{tt}},\qquad
c_b=-\frac{A_A A_{bt} + A_{tt} B_B}{A_{bt} A_{tb} - A_{bb} A_{tt}}.
\]
Again,~$A_{tt}$, $A_{bb}$~and $A_{tb}$ can be found in ref.\cite{Heinemeyer:2017gsv}.

We end up with the soft~scalar masses $m^2_\phi$ of the SSB Lagrangian. Assuming the relations
$m_i^2=c_i M_3^2$ ($i=Q,u,d,H_u,H_d$), and adding~the corrections from the gauge, the tau couplings and
$h_{\tau}$, we~get
\begin{equation}
\begin{split}
c_Q=-\frac{c_{Q{\rm Num}}}{D_m},\quad
c_u=-\frac 13\frac{c_{u{\rm Num}}}{D_m},\quad
c_d=-\frac{c_{d{\rm Num}}}{D_m},\quad
c_{H_u}=-\frac 23\frac{c_{H_u{\rm Num}}}{D_m},\quad
c_{H_d}=-\frac{c_{H_d{\rm Num}}}{D_m},
\end{split}
\end{equation}
where~$D_m$,~$c_{Q{\rm Num}}$, $c_{u{\rm Num}}$, $c_{d{\rm Num}}$, $c_{H_u{\rm Num}}$,~$c_{H_d{\rm Num}}$  and~the~complete~analysis are~again given~in~ref.~\cite{Heinemeyer:2017gsv}. These values do not obey any soft scalar mass sum rule.

If only the reduced~system was used, i.e.\ the strong, top and bottom Yukawa couplings as well as the $h_t$ and $h_b$,~the coefficients turn to be
\[
c_Q=c_u=c_d=\frac 23,\quad c_{H_u}=c_{H_d}=-1/3,
\]
which clearly~obey the sum rules
\begin{equation}
\frac{m_Q^2+m_u^2+m_{H_u}^2}{M_3^2}=c_Q+c_u+c_{H_u}=1,\qquad
\frac{m_Q^2+m_d^2+m_{H_d}^2}{M_3^2}=c_Q+c_d+c_{H_d}=1.
\end{equation}

There is an essential~point for the gaugino masses that should be mentioned.
The application of the~Hisano-Shifman relation, Eq.~(\ref{M-M0}), is made for each gaugino mass
as a boundary condition~with unified gauge coupling at $M_{\mathrm{GUT}}$.
Then, at one-loop level,~the gaugino mass depends on the one-loop coefficient of the corresponding
$\beta$-function and an~arbitrary mass $M_0$,~~ $M_i=b_iM_0$.
This fact permits, with~a suitable choice of $M_0$, to have the gluino mass equal to the unified gaugino mass,
while the gauginos~masses of the other two gauge groups are given by the gluino mass multiplied by the
ratio of the appropriate~one-loop $\beta$ coefficient.

\begin{figure}[htb!]
\begin{center}
\includegraphics[width=0.4\textwidth]{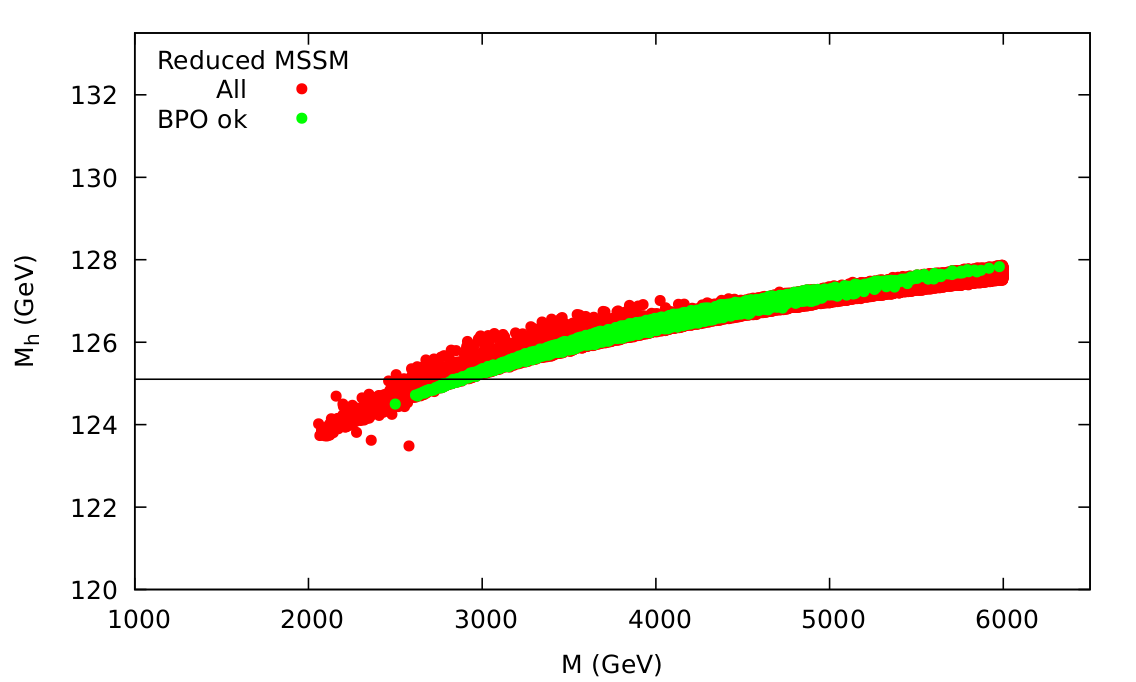}
\includegraphics[width=0.4\textwidth]{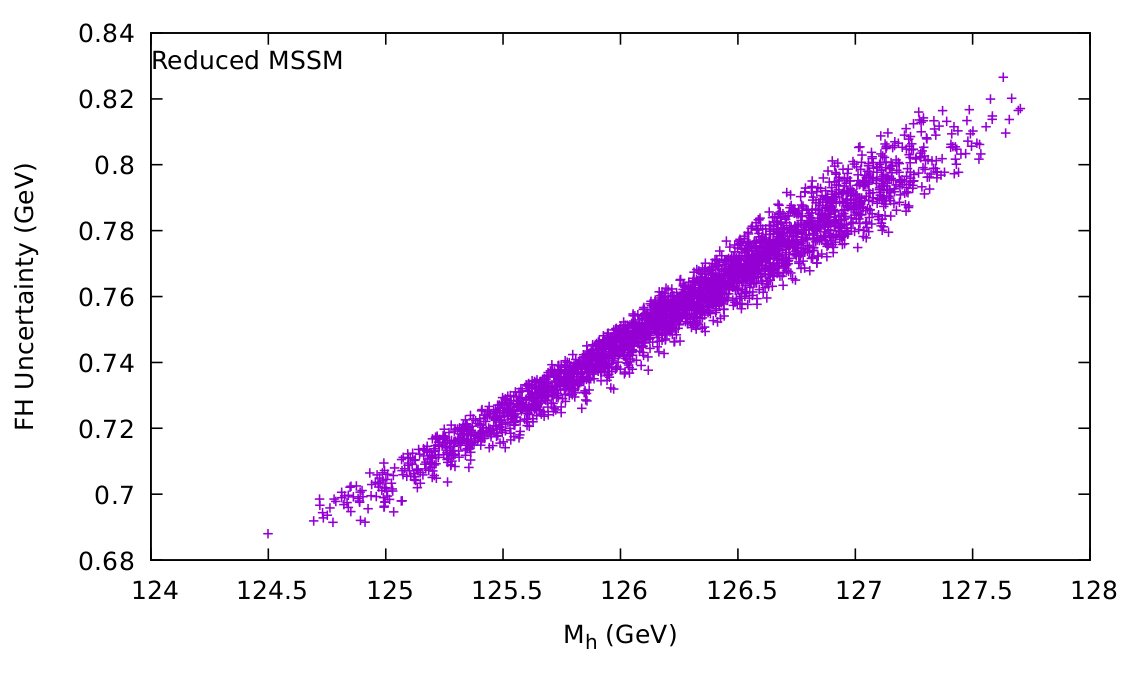}
\caption{Left: The lightest Higgs~boson mass, $\Mh$ in the Reduced MSSM. The green points is the full model
prediction. Right: the lightest~Higgs mass theoretical uncertainty \cite{Bahl:2019hmm}.}
\label{fig:rmssmhiggsvsM}
\end{center}
%\vspace{-1em}
\end{figure}

For our analysis we~choose~the unification scale to apply the~corrections
to all these RGI~relations. The full~discussion on the selection of the free parameters of the model can be found~in \cite{Heinemeyer:2020ftk}.
In total, we~vary $\rho_\tau$, $\rho_{h_\tau}$, $M$ and~$\mu$. This results in the scattered points of the next figure.

The model's~predictions for the bottom and top mass lie within 2$\sigma$ of \refeq{mtmbexp}.
The  scatter plot of the light Higgs boson mass $\Mh$ is shown in
\reffi{fig:rmssmhiggsvsM} (left),while the~theory
uncertainty given in~\reffi{fig:rmssmhiggsvsM} (right) has dropped below 1 GeV.
The Higgs mass~predicted by the model lies perfectly in the experimentally measured range.

The $\Mh$ limits~set a limit on the low-energy
supersymmetric~masses, which we briefly discuss.
The three selected~benchmarks correspond to \htb{\DRbar\ }
pseudoscalar Higgs boson masses above $1900$~GeV, $1950$~GeV and
$2000$~GeV respectively. The input of {\tt SPheno} 4.0.4
\cite{Porod:2003um,Porod:2011nf} can be found in \refta{tab:mssminput}
(notation as in \refse{sec:minimalsu5}).

%%%%%%%%%%%%%%%%%%%% T A B L E %%%%%%%%%%%%%%%%%%%%%%%%%%%%%%%%%%%%%%%%%%%%%%%
\begin{table}[htb!]
\renewcommand{\arraystretch}{1.3}
\centering\small
\begin{tabular}{|c|rrrrrrrrrr|}
\hline
  & $M_1$ & $M_2$ & $M_3$ & $|\mu|$ &
  $b~~~$ & $A_u$ & $A_d$ & $A_e$ & $\tan{\beta}$ & $m_{Q_{1,2}}^2$  \\
\hline
RMSSM-1  & 3711 & 1014 & 7109 & 4897 & $284^2$ & 5274 & 5750 & 20 & 44.9 & $5985^2$  \\
RMSSM-2  & 3792 & 1035 & 7249 & 4983 & $294^2$ & 5381 & 5871 & 557 & 44.6 & $6103^2$  \\
RMSSM-3  & 3829 & 1045 & 7313 & 5012 & $298^2$ & 5427 & 5942 & 420 & 45.3 & $6161^2$ \\
\hline
  & $m_{Q_{3}}^2$ &
  $m_{L_{1,2}}^2$ & $m_{L_{3}}^2$ & $m_{\overline{u}_{1,2}}^2$ & $m_{\overline{u}_{3}}^2$ & $m_{\overline{d}_{1,2}}^2$ & $m_{\overline{d}_{3}}^2$ & $m_{\overline{e}_{1,2}}^2$  & $m_{\overline{e}_{3}}^2$ & \\
\hline
RMSSM-1  & $5545^2$ & $2106^2$ & $2069^2$ & $6277^2$ & $5386^2$ & $5989^2$ & $5114^2$ & $3051^2$  & $4491^2$ &  \\
RMSSM-2  & $5656^2$ & $2122^2$ & $2290^2$ & $6385^2$ & $5476^2$ & $6110^2$ & $5219^2$ & $3153^2$  & $4181^2$ &  \\
RMSSM-3  & $5708^2$ & $2106^2$ & $2279^2$ & $6427^2$ & $5506^2$ & $6172^2$ & $5269^2$ & $3229^2$  & $3504^2$ &  \\
\hline
\end{tabular}

\caption{
Reduced MSSM  predictions that are used as input to {\tt SPheno}
Mass parameters are in~$\gev$ and rounded to $1 \gev$.}
\label{tab:mssminput}
\renewcommand{\arraystretch}{1.0}
\end{table}
%%%%%%%%%%%%%%%%%%%% T A B L E %%%%%%%%%%%%%%%%%%%%%%%%%%%%%%%%%%%%%%%%

Table~\ref{tab:mssmspheno} shows the resulting masses of Higgs bosons
and some of the lightest SUSY particles. The lightest neutralino (LSP) is Wino-like, as imposed by the~Hisano-Shifman relation, Eq.(\ref{M-M0}), and thus the CDM relic~density is below the
boundaries of \refeq{cdmexp}.~This renders this model~viable if
\refeq{cdmexp} is applied only as an upper limit and~additional sources
of CDM~are allowed. An additional DM component could be, e.g., a SUSY
axion~\cite{Bae:2013bva},
which would then bring the total DM density into agreement with the
Planck measurement of $\Omega_{\rm CDM} h^2$. This~is in contrast to the other three models
discussed above.

In addition, there is one more point that should be stressed. We find $\MA \lsim 1.5 \tev$ (for large values of
$\tb$ as in the other models), values substantially lower than in the
previously considered models. This can be understood as follows.
In this model, we have direct relations between the soft scalar masses and the
unified gaugino mass, which receive corrections from the two gauge couplings $g_1$ and $g_2$ and the Yukawa coupling of the $\tau$ lepton. As mentioned above, in the absence of these corrections the relations obey the soft scalar mass sum rule. However, unlike all the previous models, these corrections make the sum rule only approximate. Thus, these unique boundary conditions
result in very low values for the masses of the heavy
Higgs bosons (even compared to the minimal $SU(5)$ case presented
above, which also exhibits direct relations which however obey the sum rule).
A relatively light spectrum is also favored by the prediction for
the light CP-even Higgs boson mass, 
which turns out to be relatively high in this model and does not allow us to consider heavier spectra.
Thus, in this model, contrary 
to the models analyzed before,  because of the
large $\tan\beta\sim 45$ found here, the \textit{physical} mass of
the pseudoscalar Higgs boson, $M_A$, is excluded by the searches $H/A
\rightarrow \tau \tau$ at ATLAS with 139/fb \cite{Aad:2020zxo} for all
three benchmarks. One could try considering a heavier spectrum, in
which we would have $M_A\gtrsim 1900\gev$, but in that case the light
Higgs mass would be well above its acceptable region. Particularly, it would be above 128 GeV, a value that is clearly excluded, especially given the improved (much smaller) uncertainty calculated by the new {\tt FeynHiggs} code). Thus, the
current version of this model has been ruled out
experimentally. Consequently, we do not show any SUSY or Higgs
production cross sections.

\begin{center}
\begin{table}[ht]
\begin{center}
\small
\begin{tabular}{|l|r|r|r|r|r|r|r|r|r|r|}
\hline
  & $M_{H}$ & $M_A$ & $M_{H^{\pm}}$ & $M_{\tilde{g}}$ & $M_{\tilde{\chi}^0_1}$ & $M_{\tilde{\chi}^0_2}$ & $M_{\tilde{\chi}^0_3}$  & $M_{\tilde{\chi}^0_4}$ &  $M_{\tilde{\chi}_1^\pm}$ & $M_{\tilde{\chi}_2^\pm}$  \\\hline
RMSSM-1 & 1.393 & 1.393 & 1.387 & 7.253 & 1.075 & 3.662 & 4.889 & 4.891 & 1.075 & 4.890  \\\hline
RMSSM-2 & 1.417 & 1.417 & 1.414 & 7.394 & 1.098 & 3.741 & 4.975 & 4.976 & 1.098 & 4.976  \\\hline
RMSSM-3 & 1.491 & 1.491 & 1.492 & 7.459 & 1.109 & 3.776 & 5.003 & 5.004 & 1.108 & 5.004  \\\hline
 & $M_{\tilde{e}_{1,2}}$ & $M_{\tilde{\nu}_{1,2}}$ & $M_{\tilde{\tau}}$ & $M_{\tilde{\nu}_{\tau}}$ & $M_{\tilde{d}_{1,2}}$ & $M_{\tilde{u}_{1,2}}$ & $M_{\tilde{b}_{1}}$ & $M_{\tilde{b}_{2}}$ & $M_{\tilde{t}_{1}}$ & $M_{\tilde{t}_{2}}$ \\\hline
RMSSM-1 & 2.124 & 2.123 & 2.078 & 2.079 & 6.189 & 6.202 & 5.307 & 5.715 & 5.509 & 5.731 \\\hline
RMSSM-2 & 2.297 & 2.139 & 2.140 & 2.139 & 6.314 & 6.324 & 5.414 & 5.828 & 5.602 & 5.842 \\\hline
RMSSM-3 & 2.280 & 2.123 & 2.125 & 2.123 & 6.376 & 6.382 & 5.465 & 5.881 & 5.635 & 5.894 \\\hline
\end{tabular}
\caption{Masses for each benchmark of the Reduced MSSM (in TeV).}\label{tab:mssmspheno}
\end{center}
\end{table}
\end{center}

\section{Conclusions}

 The reduction of couplings scheme consists in searching for RGE
relations among parameters of a renormalizable  theory  that hold to
all orders in perturbation theory. In certain $N=1$ theories such a
reduction of couplings indeed appears to be theoretically realised and
therefore it developed to a powerful tool able to reduce the parameters
and increase the predictivity of these theories. In the present paper
first we briefly reviewed the ideas concerning the reduction of couplings
of renormalizable theories and the theoretical methods which have been
developed to confront the problem.  Then we turned to the question of testing experimentally the idea of 
reduction of couplings. Four
specific models, namely the Reduced Minimal $N = 1$ $SU(5)$, the all-loop
Finite $N = 1$ $SU(5)$, the two-loop Finite $N = 1$ $SU(3)^3$ and the Reduced
MSSM, have been considered for which new results have been obtained using the updated Higgs-boson mass calculation of
{\tt FeynHiggs}. In each case benchmark points in the low-mass regions have been chosen 
for which  the {\tt SPheno} code has been used to calculate  the spectrum of SUSY particles 
and  their decay modes. Finally  the {\tt MadGraph} event generator was used to compute 
the  production cross sections  of relevant final states at the 14 TeV (HL-)LHC and  100 TeV FCC-hh colliders.

The first three (unified) models were found to be in comfortable
agreement with LHC measurements and searches, with the exception of the
bottom quark mass in the Reduced Minimal $SU(5)$, for which agreement with
measurements can be achieved only at the 4$\sigma$ level. In addition it was
found that all models predict relatively heavy spectra, which evade
largely the detection in the HL-LHC.
We found one noticeable exception. The reduced MSSM features a
relatively light heavy Higgs-boson mass spectrum. Together with the
relatively high value of $\tb$ this spectrum is excluded already by
current searches at ATLAS and CMS for in the $pp \to H/A \to \tau^+\tau^-$
mode. 
We also analyzed the accessibility of the SUSY and heavy Higgs
spectrum at the FCC-hh with $\sqrt{s} = 100$~TeV.
We found that the lower parts of the parameter
space will be testable at the $2\,\sig$ level, with only an even smaller
part discoverable at the $5\,\sig$ level. However, the
heavier parts of the possible SUSY spectra will remain elusive even at the
FCC-hh. One exception here is the heavy Higgs-boson sector of the
two-loop finite $N=1$ $SU(3)^3$ model, which exhibits a spectrum
where only the highest possible mass values could escape the searches at
the FCC-hh.

%%%%%%%%%%%%%%%%%%%%%%%%%%%%%%%%%%%%%%%%%%%%%%%%%%%%%%%%%%%%%%%%%%%%%%%%%%%%%%%
%%%%%%%%%%%%%%%%%%%%%%%%%%%%%%%%%%%%%%%%%%%%%%%%%%%%%%%%%%%%%%%%%%%%%%%%%%%%%%%

\subsection*{Acknowledgements}

\noindent GZ thanks the ITP of~Heidelberg, MPI Munich, CERN Department
of Theoretical Physics, IFT Madrid and MPI-AEI for
their~hospitality. The work~of SH~is supported~in part~by the~MEINCOP
Spain~under Contract~FPA2016-78022-P and and under Contract
PID2019-110058GB-C21, in~part by~the 
Spanish~Agencia Estatal de Investigaci\'{o}n (AEI), the~EU Fondo
Europeo~de Desarrollo~Regional (FEDER) 
through the~project FPA2016-78645-P, in~part by the ``Spanish Red Consolider
MultiDark'' FPA2017-90566-REDC, and in~part by the AEI~through the grant~IFT
Centro de~Excelencia Severo Ochoa~SEV-2016-0597.
The work of~MM is partly~supported by~UNAM PAPIIT through~Grant IN111518.
The work~of GP, NT and~GZ is partially supported~by the COST~action~CA16201, GZ is also partially supported by the grant DEC-2018/31/B/ST2/02283 of NSC, Poland.
GZ has been~supported within the~Excellence Initiative~funded by the German and State~Governments,
at the Institute for~Theoretical Physics, Heidelberg University~and from the Excellence Grant~Enigmass
of LAPTh. The work of JK and WK has been supported b the National Science Centre,
Poland, the HARMONIA project under contract UMO-2015/18/M/ST2/00518
(2016-2020).

%%%%%%%%%%%%%%%%%%%%%%%%%%%%%%%%%%%%%%%%%%%%%%%%%%%%%%%%%%%%%%%%%%%%%%%%%%%%%%%%%%%%%%%%%%%%%%%%%%%%%%%

\end{document}